\documentclass[acmsmall]{acmart}
\usepackage{amsmath}
\usepackage{amsthm}
\usepackage{dsfont}
\usepackage{graphicx}
\usepackage{hyperref}
\usepackage[capitalize]{cleveref}
\usepackage{makecell}

\usepackage{pdfrender}
\usepackage{hhline}

\newtheorem{theorem}{Theorem}[section]
\newtheorem{lemma}[theorem]{Lemma}
\newtheorem{definition}{Definition}[section]
\newtheorem{corollary}{Corollary}[section]

\makeatletter
\newtheorem*{rep@theorem}{\rep@title}
\newcommand{\newreptheorem}[2]{%
\newenvironment{rep#1}[1]{%
 \def\rep@title{#2 \ref{##1}}%
 \begin{rep@theorem}}%
 {\end{rep@theorem}}}
\makeatother

\newreptheorem{theorem}{Theorem}
\newreptheorem{lemma}{Lemma}

\mathchardef\hy="2D

\newcommand*{\boldcheckmark}{%
  \textpdfrender{
    TextRenderingMode=FillStroke,
    LineWidth=.5pt, 
  }{\checkmark}%
}

\newcommand{\undone}{\textsc{undone}}
\newcommand{\game}{\textsc{game}}
\newcommand{\service}{\textsc{service}}
\newcommand{\need}{\textsc{need}}
\newcommand{\remain}{\textsc{remain}}
\newcommand{\sumsc}{\textsc{sum}}

\setcopyright{acmlicensed}
\acmJournal{POMACS}
\acmYear{2022} \acmVolume{6} \acmNumber{3} \acmArticle{51} \acmMonth{12} \acmPrice{}
\acmDOI{10.1145/3570612}

\received{August 2022}
\received[revised]{October 2022}
\received[accepted]{November 2022}

\title{Optimal Scheduling in the Multiserver-job Model under Heavy Traffic}
\author{Isaac Grosof}
\email{igrosof@cs.cmu.edu}
\orcid{0000-0001-6205-8652}
\affiliation{%
    \institution{Carnegie Mellon University}
    \department{Computer Science Department}
    \city{Pittsburgh}
    \state{PA}
    \country{USA}
}
\author{Ziv Scully}
\email{zscully@cs.cmu.edu}
\orcid{0000-0002-8547-1068}
\affiliation{%
    \institution{Simons Institute, UC Berkeley \& Carnegie Mellon University \& Cornell University}
    \department{Department of Electrical Engineering and Computer Sciences}
    \city{Berkeley}
    \state{CA}
    \country{USA}
}
\author{Mor Harchol-Balter}
\email{harchol@cs.cmu.edu}
\orcid{0000-0003-1721-6759}
\affiliation{%
    \institution{Carnegie Mellon University}
    \department{Computer Science Department}
    \city{Pittsburgh}
    \state{PA}
    \country{USA}
}
\author{Alan Scheller-Wolf}
\email{awolf@andrew.cmu.edu}
\orcid{0000-0001-6871-2360}
\affiliation{%
    \institution{Carnegie Mellon University}
    \department{Tepper School of Business}
    \city{Pittsburgh}
    \state{PA}
    \country{USA}
}



\begin{document}
\begin{abstract}
Multiserver-job systems, where jobs require concurrent service
at many servers, occur widely in practice.
Essentially all of the theoretical work on multiserver-job systems
focuses on maximizing utilization,
with almost nothing known about mean response time.
In simpler settings,
such as various known-size single-server-job settings,
minimizing mean response time is merely a matter of prioritizing small jobs.
However, for the multiserver-job system,
prioritizing small jobs is not enough,
because we must also ensure servers are not unnecessarily left idle.
Thus, minimizing mean response time
requires prioritizing small jobs while simultaneously maximizing throughput.
Our question is how to achieve these joint objectives.

We devise the ServerFilling-SRPT scheduling policy,
which is the first policy to minimize mean response time in the multiserver-job model
in the heavy traffic limit.
In addition to proving this heavy-traffic result,
we present empirical evidence that ServerFilling-SRPT
outperforms all existing scheduling policies for all loads,
with improvements by orders of magnitude at higher loads.

Because ServerFilling-SRPT requires knowing job sizes,
we also define the ServerFilling-Gittins policy,
which is optimal when sizes are unknown or partially known.
\end{abstract}
\maketitle
\section{Introduction}
\label{sec:introduction}
\subsection{The multiserver-job model}

Traditional multiserver queueing theory focuses on models, such as the M/G/$k$,
where every job occupies exactly one server.
For decades, these models remained popular because they captured the behavior of
computing systems, while being amenable to theoretical analysis.
However, such one-server-per-job models
are no longer representative of many modern computing systems.

\begin{figure}
    \centering
    \includegraphics[width=0.6\textwidth]{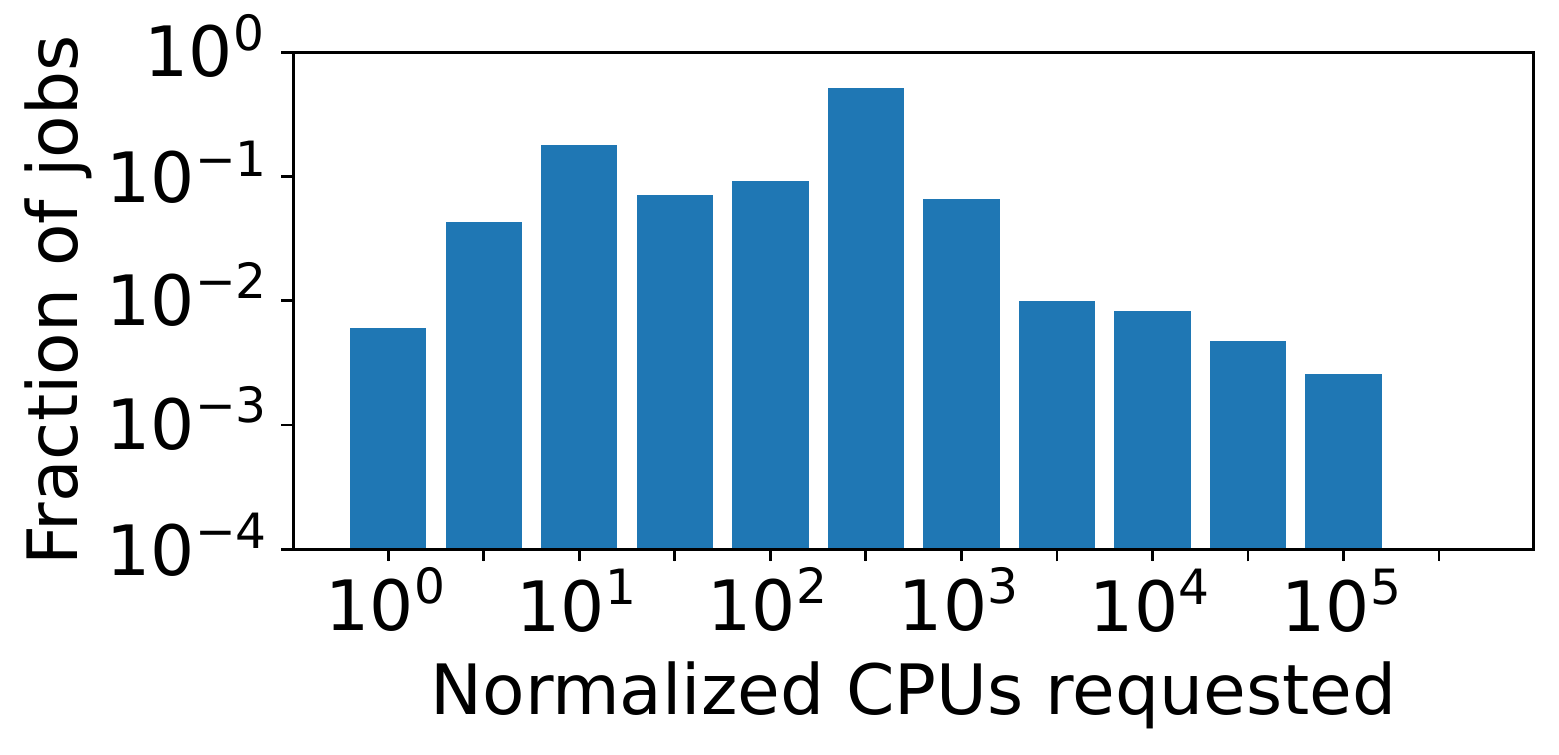}
    \caption{The distribution of number of CPUs requested by jobs in Google's recently published Borg trace \cite{tirmazi_2020}. Number of CPUs is normalized so that the smallest job in the trace uses one normalized CPU.}
    \label{fig:dist}
\end{figure}

Consider today's large-scale computing centers,
such as the those of Google, Amazon and Microsoft.
While the \emph{servers} in these data centers still resemble the \emph{servers}
in traditional models such as the M/G/$k$,
the \emph{jobs} have changed:
Each job now requires many servers, which it holds simultaneously.
While some jobs require few servers, other jobs require many more servers.
For instance, in \cref{fig:dist}, we show the distribution of the number of CPUs
requested by jobs in Google's recently published trace of its ``Borg'' computation cluster
\cite{tirmazi_2020,grosof_wcfs_2021}.
The distribution is highly variable, with jobs requesting anywhere from 1 to 100,000 normalized CPUs.
Throughout this paper, we will focus on this ``multiserver-job model'' (MSJ),
by which we refer to the common situation in modern systems where each job concurrently occupies
a fixed number of servers (typically more than one),
throughout its time in service.

The multiserver-job model is fundamentally different from the one-server-per-job model.
In the one-server-per-job model,
any work-conserving scheduling policy such as First-Come First-Served (FCFS) can
achieve full server utilization.
In the multiserver-job model,
a na\"ive scheduling policy such as FCFS will waste more servers than necessary.
As a result, server utilization and system stability are dependent on the scheduling policy
in the multiserver-job model.
While finding throughput-optimal scheduling policies is a challenge,
several such policies are known, including MaxWeight \cite{maguluri_2012},
Randomized Timers \cite{ghaderi_randomized_2016,psychas_randomized_2018},
and ServerFilling \cite{grosof_wcfs_2021}.
However, none of these policies give consideration to optimizing mean response time;
each policy solely focuses on optimizing throughput.
In fact, the empirical mean response time of such policies
can be very poor \cite{ghaderi_randomized_2016},
motivating our goal of finding throughput-optimal policies
which moreover minimize mean response time.

\subsection{The challenges of minimizing MSJ mean response time}

In the M/G/$k$ setting, where each job requires a single server,
it was recently proven that the SRPT-$k$ (Shortest Remaining Processing Time-$k$)
scheduling policy minimizes mean response time in the heavy-traffic limit \cite{grosof_srpt_2018}.
SRPT-$k$ is a very simple policy: serve the $k$ jobs of least remaining duration (service time).

Unfortunately, in the multiserver-job system,
trying to simply adapt the SRPT-$k$ policy
does not result in an optimal policy
for two reasons:
\begin{itemize}
    \item Prioritizing by remaining job duration
    is not the right way to minimize mean response time.
    We will show that an optimal policy must prioritize by remaining \emph{size},
    which we define to be proportional to the product of
    a job's duration and its \emph{server need},
    the number of servers the job requires.
    We define these terms in more detail in \cref{sec:setting}.
    \item Even with this concept of size,
    a prerequisite for minimizing mean response time
    in the heavy-traffic limit
    is throughput-optimality, which requires a policy to efficiently utilize
    all of the servers whenever possible.
    Unfortunately, greedily prioritizing the job of least remaining size,
    as in SRPT-k, is not throughput optimal.
    Our policy must be throughput-optimal, while \emph{also} prioritizing small jobs.
\end{itemize}

We therefore ask:
\begin{quote}
\textit{What scheduling policy for the multiserver-job model
    should we use to \emph{minimize mean response time}
    in the heavy-traffic limit?}
\end{quote}

By ``heavy-traffic'' we mean as load $\rho \to 1$,
while the number of servers, $k$, stays fixed.
The precise definition of load $\rho$
and the heavy-traffic limit will be explained in detail in \cref{sec:setting}.

\subsection{ServerFilling-SRPT and ServerFilling-Gittins}
To answer this question, we introduce the ServerFilling-SRPT scheduling policy,
the first scheduling policy to minimize mean response time in the multiserver-job model
in the heavy traffic limit.

ServerFilling-SRPT is defined in the setting where $k$ is a power of 2,
and all server needs are powers of 2.
This setting is commonly seen in practice in supercomputing
and other highly-parallel computing settings
\cite{cirne_model_2001,downey_using_1997}.

To define ServerFilling-SRPT,
imagine all jobs are ordered by their remaining size.
Select the smallest initial subset $M$ of this sequence
such that the jobs in $M$ collectively require at least $k$ servers.
Finally, place jobs from $M$ into service
in order of largest server need.
This procedure is performed preemptively,
whenever a job arrives or completes.
As we show in \cref{sec:server-filling-defn},
whenever jobs with total server need at least $k$
are present in the system,
this procedure will fill all $k$ servers.
We use this property to prove in \cref{sec:sfs}
that ServerFilling-SRPT minimizes mean response time in the heavy-traffic limit.

ServerFilling-SRPT requires the scheduler to know job durations, and hence sizes, in advance.
Sometimes the scheduler does not have duration information.
In the M/G/1 setting,
when job sizes are unknown, the Gittins policy \cite{gittins_multi_2011}
is known to achieve optimal mean response time.
We therefore introduce the ServerFilling-Gittins policy in \cref{sec:gittins-outline}.
We prove similar heavy-traffic optimality results for ServerFilling-Gittins.

\subsection{A generalization: DivisorFilling-SRPT and DivisorFilling-Gittins}

While ServerFilling-SRPT requires that the server needs are powers of 2,
we have developed a more general scheduling policy
which requires only that the server needs all divide $k$.
We call this generalization DivisorFilling-SRPT.
The DivisorFilling-SRPT policy is more complex than ServerFilling-SRPT,
and hence we defer its discussion to \cref{app:divisor-filling}.
In \cref{app:divisor-filling},
we define both DivisorFilling-SRPT and DivisorFilling-Gittins.
We then show that all of our results about ServerFilling-SRPT and ServerFilling-Gittins
hold for DivisorFilling-SRPT and DivisorFilling-Gittins.

\subsection{A Novel Proof Technique: MIAOW}

In recent years, there have been a plethora of proof techniques developed
to handle the analysis of multiserver systems.
These include:
\begin{itemize}
    \item Multiserver tagged job analysis
    \cite{grosof_srpt_2018,grosof_guardrails_2019,scully_optimal_2021},
    \item Worst-case work gap
    \cite{grosof_srpt_2018,grosof_guardrails_2019,scully_optimal_2021},
    \item WINE (Work Integral Number Equality)
    \cite{scully_gittins_2020,scully_wine_2021},
    \item Work Decomposition law \cite{scully_gittins_2020}.
\end{itemize}
Unfortunately,
none of these techniques suffice to handle the analysis of ServerFilling-SRPT
and DivisorFilling-SRPT.
As we discuss in \cref{sec:miaow}, the analysis of ServerFilling-SRPT
requires bounding the \emph{waste} relative to
a resource-pooled single-server SRPT system,
where waste is the expected product of work and unused system capacity.
In order to analyze waste,
we introduce a new technique called MIAOW,
Multiplicative Interval Analysis of Waste.
MIAOW subdivides jobs into multiplicative intervals based on their remaining sizes,
and bounds the waste in each interval.
\subsection{Comparison with other policies}

In \cref{tbl:comparison}, we compare our ServerFilling-SRPT and DivisorFilling-SRPT policies
and our asymptotic optimality results
with prior work in the multiserver-job setting.
Prior work broadly falls into two categories:
theoretical results focusing on throughput-optimality,
and good heuristic policies.
Our result is the first to theoretically study the problem of minimizing mean response time.

\begin{table}
\centering
\begin{tabular}{|p{50mm}|c|c|c|c|}
\hline
{} & \multicolumn{2}{c|}{Maximize throughput} & \multicolumn{2}{c|}{Minimize mean response time} \\
\hline
Policies & Attempted & Proven & Attempted & Proven \\
\hhline{|=|=|=|=|=|}
MaxWeight \cite{maguluri_2012} & \checkmark & \checkmark & {} & {} \\
\hline
Randomized Timers \cite{ghaderi_randomized_2016,psychas_randomized_2018}
& \checkmark & \checkmark & {} & {} \\
\hline
ServerFilling \cite{grosof_wcfs_2021}
& \checkmark & \checkmark & {} & {} \\
\hline
FCFS \cite{feitelson_toward,jones_scheduling,rumyantsev_2017}
& {} & {} & {} & {} \\
\hline
Simple backfilling heuristics:
FirstFit, BestFit, etc. \cite{wang_application_2009,jones_scheduling}
& \checkmark & {} & {} & {} \\
\hline
Size-aided backfilling:
EASY, conservative, dynamic, etc.
\cite{jones_scheduling,carastan_one_2019}
& \checkmark & {} & {} & {} \\
\hline
Size-based heuristics: GreedySRPT, FirstFitSRPT, etc. \cite{carastan_one_2019}
& {} & {} & \checkmark\footnotemark  & {} \\
\hline
Size \& learning heuristics \cite{guo_optimal_2018}
& \checkmark & {} & \checkmark & {} \\
\hline
\textbf{ServerFilling-SRPT} (\cref{sec:sfs}) \&
\textbf{DivisorFilling-SRPT} (App.~\ref{app:divisor-filling})
& \boldcheckmark & \boldcheckmark & \boldcheckmark & \boldcheckmark \\
\hline
\end{tabular}
\caption{Comparison of multiserver-job scheduling policies}
\label{tbl:comparison}
\end{table}

\begin{figure}
    \centering
    \includegraphics[width=0.8\textwidth]{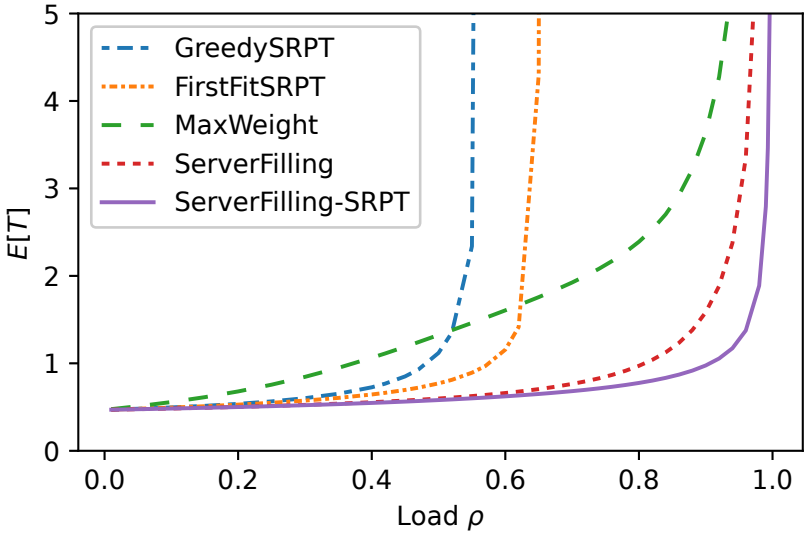}
    \caption{Simulated mean response time $E[T]$ as a function of load $\rho$
    in a multiserver-job setting with $k=8$ total servers.
    Server need is sampled uniformly from $\{1, 2, 4, 8\}$.
    Size is exponentially distributed,
    independent of the number of servers required. Policies defined in \cref{sec:empirical}.
    Simulations use $10^7$ arrivals. Loads $\rho \in [0, 0.999]$ simulated.}
    \label{fig:intro-mean}
\end{figure}

\cref{fig:intro-mean}
compares the mean response time of ServerFilling-SRPT to that of
prior throughput-optimal policies, as well as na\"ive size-based heuristic policies.
These selected policies are representative of the empirical
behavior of a wide variety of prior policies:
Some of the policies shown have SRPT-like behavior,
some policies are throughput-optimal,
but only our ServerFilling-SRPT policy achieves both.
Correspondingly, in this simulation and others we have performed,
ServerFilling-SRPT has the best mean response time at all loads $\rho$,
often by huge margins.
\footnotetext{
Because these heuristics are not throughput optimal,
they are only competitive for mean response time at low to moderate load $\rho$.
}

\subsection{Summary of our contributions and outline}
\begin{itemize}
    \item In \cref{sec:setting},
    we introduce the ServerFilling-SRPT scheduling policy.
    \item In \cref{sec:sfs},
    we bound the mean response time of ServerFilling-SRPT.
    We introduce MIAOW, a new technique for bounding the total ``relevant'' work in the system.
    Using that bound, we prove that ServerFilling-SRPT
    has asymptotically optimal mean response time as load $\rho \to 1$.
    \item In \cref{sec:gittins-outline},
    we introduce the ServerFilling-Gittins scheduling policy,
    in the setting of unknown or partially-known job sizes and durations.
    We prove a similar bound and asymptotic optimality result for ServerFilling-Gittins.
    \item In \cref{sec:empirical},
    we empirically evaluate ServerFilling-SRPT using simulation,
    showing that it outperforms prior policies on realistic distributions over a variety of loads,
    not just the $\rho \to 1$ limit.
\end{itemize}
All of our results for ServerFilling-SRPT and ServerFilling-Gittins
also extend to DivisorFilling-SRPT and DivisorFilling-Gittins.
\section{Prior work}
There are no prior optimality or asymptotic optimality results for mean response time
in the multiserver-job system.
The most similar system where such results have been proven is the M/G/$k$,
a multiserver system with single-server jobs,
and those results build off of classical results in the M/G/1.

\subsection{Single-server-job models (one server per job)}

In the single-server setting,
the Shortest Remaining Processing Time policy (SRPT),
which prioritizes the job of least remaining size,
has been proven to minimize mean response time in the known-size M/G/1,
as well as the worst-case single-server system \cite{schrage_srpt_1966,schrage_proof_1968}.
Note that in the single-server setting,
a job's size is simply its duration.
In the unknown- and partially-known-size settings,
the Gittins policy is known to minimize mean response time in the M/G/1
\cite{gittins_multi_2011,scully_gittins_2021}. 

In the M/G/$k$, where jobs require a single server,
\cite{grosof_srpt_2018}
proves that the SRPT-$k$ scheduling policy,
the natural analogue of SRPT in the M/G/$k$,
asymptotically minimizes mean response time in the known-size M/G/$k$ in the heavy-traffic limit.
There, as in this paper, load $\rho$ is defined as the long-term average fraction of busy servers;
$\rho \to 1$ is the heavy-traffic limit.
Specifically, the paper shows that
\begin{align*}
    \lim_{\rho \to 1} \frac{E[T^{SRPT \hy k}]}{E[T^{OPT \hy k}]}
    =1.
\end{align*}
This is proven despite the fact that the optimal policy OPT-$k$ is unknown.

In the unknown job size setting, similar asymptotic optimality results
for mean response time have been proven
for the Gittins-$k$ policy \cite{scully_gittins_2020}
and a monotonic variant thereof
\cite{scully_optimal_2021}.
Moreover, for the Gittins-$k$ policy, these results generalize
to the partially-known job size setting, such as a setting with imperfect job size estimates.

\subsection{Multiserver-job model (many servers per job)}
Theoretical results in the multiserver-job model are limited.
The \emph{blocking} model, where arriving jobs either immediately receive service or are dropped,
has received significant attention, with many strong results
\cite{arthurs_sizing_1979,whitt_blocking_1985,vandijk_blocking_1989,tikhonenko_generalized_2005}
such as the exact steady state distribution.
However, without any queue these models don't fit most real computing systems well.
In the \emph{queueing} MSJ model,
which we focus on,
results are much more limited \cite{harchol_multiserver_2022}.
The best-studied scheduling policy is the first-come first-served (FCFS) policy.
Stability region results for FCFS are known in several limited settings
\cite{rumyantsev_2017,rumyantsevi_three_2022},
and steady state results are only known in the case of two servers
\cite{kim_mms_1979,brill_queues_1984,fillippopoulos_mm2}.

Recently, the Work Conserving Finite Skip (WCFS) framework has been used to
analytically characterize response time
under the ServerFilling and DivisorFilling scheduling policies
\cite{grosof_wcfs_2021},
both of which serve jobs in near-FCFS order.
We modify the ServerFilling and DivisorFilling policies
to prioritize jobs of shortest remaining size (SRPT).
We then use a novel proof technique called MIAOW
to demonstrate that ServerFilling-SRPT and DivisorFilling-SRPT
achieve asymptotically similar mean response time to SRPT in an analogous M/G/1 setting.

There has also been work in the \emph{scaling} multiserver-job model,
where one analyzes a sequence of multiserver-job systems
with jointly increasing arrival rate, number of servers,
and server needs \cite{wang_zero_2021,hong_sharp_2022}.
The regimes investigated include multiserver-job analogues
of the Halfin-Whitt regime.
Our results complement these, as we
study a system with a fixed number of servers $k$ in the heavy-traffic limit.
\subsection{Supercomputing}
Supercomputing centers are one of the originators of the multiserver-job model:
Supercomputing jobs closely resemble 
the jobs in the multiserver-job model.
Jobs commonly demand anywhere from one core to thousands of concurrent cores
\cite{msi_queues,vizino_batch}.
Unfortunately, all of the papers in this area
focus on simulation or empirical results,
rather than analytical results \cite{feitelson_toward,jones_scheduling,feitelson_parallel,armstrong_scheduling,tang_reducing,etsion_short,tang_adaptive}.
These papers study a variety of scheduling policies,
such as FCFS, various backfilling policies,
and other more novel policies.
Backfilling policies considered include simpler, no-duration-information policies such as
FirstFit and BestFit \cite{wang_application_2009,jones_scheduling},
as well as more complex, duration-information-based policies such as EASY backfilling \cite{srinivasan_characterization_2002},
conservative backfilling \cite{srinivasan_characterization_2002},
Smallest Area%
\footnote{The term ``area'' used in \cite{carastan_one_2019} is equivalent to our ``size''.}
First-backfilling \cite{carastan_one_2019},
Dynamic Backfilling \cite{jones_scheduling}, and many more.

Often the primary goal of these papers is achieving high utilization,
with secondary goals including
minimizing mean response time and ensuring fairness between different types of jobs.
However, their settings are sometimes more restrictive than our setting:
preemption may be either limited or impossible.
When preemption is impossible, maximum utilization is lower, often around $\rho = 70\%$,
and mean response times are often high near the utilization threshold.

Our scheduling policies, such as ServerFilling-SRPT,
can only be defined for the subset of settings
where preemption is possible,
and our policies leverage preemption to achieve much stronger results in those settings.
\subsection{Virtual Machine Scheduling}
In the field of cloud computing, the Virtual Machine (VM) scheduling problem
is essentially a multi-resource generalization of the multiserver-job model.
In this model, rather than a single requirement like server need,
each job requires concurrent utilization of several different limited resources,
such as RAM, CPU, GPU, network bandwidth, etc.
Of course, any results in this more general setting also apply to the multiserver-job setting.
In the VM scheduling literature, papers typically focus on finding a throughput-optimal policy.
Two major categories of such policies are the preemptive MaxWeight \cite{maguluri_2012}
and non-preemptive Randomized Timers \cite{ghaderi_randomized_2016,psychas_randomized_2018}
scheduling frameworks.

These papers focus entirely on achieving throughput optimality,
and the mean response time of the resulting policies can be poor,
as several of the above papers note.
Work on optimal mean response time in the VM scheduling literature
has been limited to heuristic policies and empirical evaluation \cite{guo_optimal_2018}.

\section{Setting}
\label{sec:setting}
\subsection{Multiserver-job Model}
The multiserver-job (MSJ) model is a multiserver queueing model
where each job requires a fixed number of servers concurrently over its entire time in service.
The jobs are therefore called ``multiserver jobs.''

A job $j$ has two requirements: A server need $k_j$ and a service duration $d_j$.
These requirements are sampled i.i.d. from some joint distribution
with random variables $(K, D)$.
Note that $K$ and $D$ can be correlated.
A job's server need $k_j$ is at most the total number of servers, $k$.
The total server need of the jobs in service at any time must sum to at most $k$.
The job $j$ will complete after $d_j$ time in service.

We assume Poisson arrivals with rate $\lambda$,
and we assume preemption is allowed with no loss of progress.

Let a job $j$'s \emph{size} $s_j$ be defined as $k_j d_j/k$,
and likewise define the job size distribution $S = KD/k$.
Job $j$'s size can be viewed as
the area of a rectangle with height equal to the job's duration $d_j$
and width equal to $k_j/k$, the fraction of the total service capacity occupied by job $j$.
Likewise, a job's \emph{remaining size} $r_j$ is its remaining duration multiplied by $k_j/k$.
We define a job $j$'s \emph{service rate} to be $k_j/k$,
the rate at which the job's remaining size decreases during service.
We define a job's age $a_j$ to be $s_j - r_j$,
which increases at rate $k_j/k$ whenever the job is in service.

A \emph{resource-pooled} M/G/1 is defined to be a system
with a single server with the same capacity as all $k$ original servers pooled together,
and the same arrival rate $\lambda$ and job size distribution $S$ as the original MSJ system.
We allow the resource-pooled M/G/1 to divide its capacity arbitrarily among
the jobs in the system.
In particular, while jobs in the MSJ system have fixed service rates depending on their server needs,
in the resource-pooled system any combination of service rates is allowed,
decreasing remaining sizes accordingly.
Note that the resource-pooled system is strictly more flexible than the MSJ system,
so the optimal policy in the resource-pooled system
is superior to the optimal policy in the MSJ system.

Let $W(t)$ be the total work in the system at time $t$:
The sum of the remaining sizes $r_j$ of all job's in the system at time $t$.
Let $B(t)$ be the \emph{``busyness''} of the system at time $t$:
The fraction of servers that are occupied at time $t$.
Note that $B(t)$ is also the total service rate of all jobs in service at time $t$,
and so $B(t) = -\frac{d}{dt} W(t)$,
outside of arrival moments.
We also define $W$ and $B$ to be the corresponding stationary random variables.
 
Let \emph{load} $\rho = \lambda E[S]$ be the long-run average rate
at which work arrives to the system.
We assume $\rho < 1$ as a necessary condition for stability.
We will focus on settings where $\rho < 1$ is also sufficient for stability
for some feasible scheduling policy.
Note that $\rho$ is a constant
and that $\rho = E[B]$,
under any scheduling policy for which the system is stable.

Next, let us define an \emph{$r$-relevant} job, where $r$ is a remaining size threshold.
A job $j$ is $r$-relevant if $r_j \le r$.
This terminology is in reference to the tagged job analysis used in studying SRPT
in the M/G/1 and M/G/$k$ settings \cite{schrage_srpt_1966,grosof_srpt_2018};
in those settings, the service of a job with remaining size $r$ is only affected by
the presence of $r$-relevant jobs in the system.
The multiserver-job system is not as simple,
so we do not employ a tagged-job approach, but we reuse the terminology.

Correspondingly, let the $r$-relevant work $W_r(t)$ be
the total remaining size of all $r$-relevant jobs in the system at time $t$,
and let $B_r(t)$ be the fraction of servers which are serving $r$-relevant jobs at time $t$.
Define $B_r$ and $W_r$ correspondingly.
The core of our proof lies in bounding
expectations of random variables involving $B_r$ and $W_r$,
and combining these with a characterization of mean response time $E[T]$
in terms of $B_r$ and $W_r$.

Next, let us define the $r$-relevant load $\rho_r$
to be the long-run average $r$-relevant busyness of the system.
A job with size $s_j$ receives $\min(s_j, r)$
service while having remaining size $\le r$.
As a result, $\rho_r = \lambda E[\min(S, r)] = E[B_r]$.
We further divide the $r$-relevant load based
on whether the job in question has initial size $\le r$.
Let the \emph{arrival load} $\rho^A_r = \lambda E[S \mathds{1} \{S < r\}]$,
and let the \emph{recycled load} $\rho^R_r = \lambda r P(S > r)$.
Note that $\rho_r = \rho^A_r + \rho^R_r$.
Note also that $\rho_r, \rho^A_r,$ and $\rho^R_r$
are all not dependent on the policy $\pi$.

Finally, let us define an \emph{$r$-recycling moment} to be a moment when a job $j$ with initial size
$s_j > r$ reaches remaining size $r_j = r$.
Let $E_r[\cdot]$ be an expectation taken over $r$-recycling moments,
just prior to the job recycling.
\subsection{ServerFilling-SRPT}
\label{sec:server-filling-defn}
This paper considers two settings of server needs:
\begin{itemize}
    \item The ``power of two'' setting:
        $k$ is power of two, and all server needs $k_j$ are powers of two.
    \item The ``divisible'' setting:
        $k$ is general, and all server needs $k_j$ are divisors of $k$.
\end{itemize}
Corresponding to these two settings, we have two policies of interest:
ServerFilling-SRPT for the power of two setting,
and DivisorFilling-SRPT for the divisible setting.
We define ServerFilling-SRPT here,
and DivisorFilling-SRPT in \cref{app:divisor-filling}.
When writing equations throughout the paper, we abbreviate ServerFilling-SRPT
as SFS-$k$.

To implement SFS-$k$,
start by ordering jobs in increasing order of remaining size $r_j$,
breaking ties arbitrarily.
Define $j_1, j_2, \ldots$ such that
\begin{align*}
    r_{j_1} \le r_{j_2} \le \ldots.
\end{align*}
Next, consider initial subsets of this ordering: 
\begin{align*}
    \{j_1\}, \{j_1, j_2\},  \{j_1, j_2, j_3\} \ldots.
\end{align*}
We are interested in the smallest initial subset $M$
in which the total server need is at least $k$.
In other words, let $i^*$ be the smallest index such that
\begin{align*}
    \sum_{i=1}^{i^*} k_{j_i} \ge k.
\end{align*}
If there is no such index, then ServerFilling-SRPT serves all jobs in the system simultaneously.

Otherwise, ServerFilling-SRPT will serve a subset of $M = \{j_1, j_2, \ldots, j_{i^*}\}$.
Among this subset, ServerFilling-SRPT prioritizes jobs of largest server need,
placing jobs into service in descending order of server need,
until no servers remain or the next job cannot fit, breaking ties by smallest remaining size,
and further ties arbitrarily.

In the power-of-two setting,
ServerFilling-SRPT guarantees the following strong property:
At all times, either all servers are busy, or all jobs are in service.
This was proven for the ServerFilling policy \cite[Lemma~1]{grosof_wcfs_2021},
which is identical to ServerFilling-SRPT, except that jobs are ordered in arrival order,
rather than SRPT order.
For completeness, we reprove this result here:

\begin{lemma}
    \label{lem:sfs-packs}
    Under the ServerFilling-SRPT policy,
    in the power-of-two setting,
    if the total server need of jobs in the system is at least $k$ servers,
    all $k$ servers are busy.
\end{lemma}
\begin{proof}
    Recall that $M$ is a set of jobs,
    each with server need a power of two,
    which have a total server need of at least $k$.
    Label the jobs $m_1, m_2, \ldots$
    in decreasing order of server need,
    tiebroken by least remaining size.
    \begin{align*}
        k_{m_1} \ge k_{m_2} \ge \ldots
    \end{align*}
    Let $\need(z)$ represent the total server need of the first $z$ jobs in this ordering:
    \begin{align*}
        \need(z) = \sum_{i=1}^z k_{m_i}.
    \end{align*}
    The set of jobs served by ServerFilling-SRPT is an initial sequence of this server need ordering:
    $\{m_i \mid i \le \ell\}$ for some $\ell$.
    Specifically, the index $\ell$ up to which ServerFilling-SRPT serves jobs
    is the largest index $z$ such that $\need(z) \le k$.
    To prove \cref{lem:sfs-packs},
    it suffices to show that $\need(\ell) = k$.

    Note that
    $\need(0) = 0$ and $\need(|M|) \ge k$.
    As a result, $\need(z)$ must cross $k$ at some point.
    To prove that $\need(\ell) = k$, it suffices to prove that:
    \begin{align}
        \label{eq:need}
        \text{There exists no index $\ell'$ such that $\need(\ell') < k$
        and $\need(\ell' + 1) > k$.}
    \end{align}
    To prove \eqref{eq:need},
    let us define $\remain(z)$, the number of servers remaining after $z$
    jobs have been placed into service:
    \begin{align*}
        \remain(z) = k - \need(z).
    \end{align*}
    Because all server needs $k_j$ are powers of two,
    we will show that $\remain(z)$ carries an important property:
    \begin{align}
        \label{eq:remain}
        \remain(z) \text{ is divisible by } k_{m_{z+1}} \text{ for all }z.
    \end{align}
    We will use \eqref{eq:remain} to prove \eqref{eq:need}.
    We write $a|b$ to indicate that $a$ divides $b$.

    We will prove \eqref{eq:remain} by induction on $z$.
    For $z = 0$, $\remain(0) = k$.
    Because $k$ is a power of two,
    and $k_{m_1}$ is a power of two no greater than $k$,
    the base case holds.
    Next, assume that \eqref{eq:remain} holds for some index $z$,
    meaning that $k_{m_{z+1}} | \remain(z)$.
    Note that $\remain(z+1) = \remain(z) - k_{m_{z+1}}$.
    As a result, $k_{m_{z+1}} | \remain(z+1)$.
    Now, note that $k_{m_{z+2}} | k_{m_{z+1}}$,
    because both are powers of two, and $k_{m_{z+2}} \le k_{m_{z+1}}$.
    As a result, $k_{m_{z+2}}|\remain(z+1)$, completing the proof of \eqref{eq:remain}.

    Now, we are ready to prove \eqref{eq:need}.
    Assume for contradiction that such an $\ell'$ exists.
    Then $\remain(\ell') > 0$, and $\remain(\ell'+1) < 0$.
    Because $\remain(\ell'+1) = \remain(\ell') - k_{m_{\ell'+1}}$,
    we therefore know that $k_{m_{\ell'+1}} > \remain(\ell')$.
    But from \eqref{eq:remain}, we know that $k_{m_{\ell'+1}}$ divides
    $\remain(\ell')$, which is a contradiction.
\end{proof}
Note that \cref{lem:sfs-packs} remains true if the power-of-two setting is replaced by the power-of-$x$ setting, for any integer $x$.
In fact, the only condition on the server needs necessary to prove \cref{lem:sfs-packs}
is that all server needs divide $k$, and all server needs divide all larger server needs.

An important corollary of \cref{lem:sfs-packs} is a property which we call ``relevant work efficiency'':
\begin{corollary}[Relevant work efficiency]
    \label{cor:key-sfs}
    Under the ServerFilling-SRPT policy,
    in the power-of-two setting,
    if there are $k$ or more $r$-relevant jobs in the system,
    all servers are occupied by $r$-relevant jobs,
    meaning that $B_r = 1$.
\end{corollary}
\begin{proof}
Note that $|M| \le k$,
because $M$ is the smallest initial subset of the SRPT ordering
with total server need at least $k$,
and all jobs have server need at least 1.
Therefore, if there are $k$ or more $r$-relevant jobs in the system,
then all jobs in $M$ are $r$-relevant,
so ServerFilling-SRPT fills all $k$ servers with $r$-relevant jobs,
meaning that $B_r = 1$.
\end{proof}

\cref{cor:key-sfs} is the sole property of ServerFilling-SRPT that we will use to prove
our main theorems, \cref{thm:sfs-bound,thm:sfs-opt}.

DivisorFilling-SRPT in the divisible setting also satisfies the relevant work efficiency property:
If there are $k$ or more $r$-relevant jobs in the system, then $B_r = 1$,
as we discuss in \cref{app:divisor-filling}.
As a result, our main theorems, \cref{thm:sfs-bound,thm:sfs-opt}, also hold for DivisorFilling-SRPT.
\section{ServerFilling-SRPT: Asymptotically Optimal Mean Response Time}
\label{sec:sfs}
\subsection{Summary of Results and Proofs}
To prove the optimality of ServerFilling-SRPT,
we will compare ServerFilling-SRPT's mean response time
against a resource-pooled M/G/1/SRPT system with the same size distribution $S$.
Let ``SRPT-1'' denote the M/G/1/SRPT system.
Recall that SRPT-1
combines the power of all $k$ servers into a single server,
which can work on any job or any mixture of jobs.
This resource-pooled system is strictly more flexible than the multiserver-job system,
so the optimal policy in the resource-pooled system forms a lower bound
on the optimal policy in the MSJ system.
Because SRPT minimizes mean response time in the M/G/1,
SRPT-1 yields a lower bound on the optimal mean response time in the MSJ system.

We will upper bound the gap in mean response time
between ServerFilling-SRPT
and SRPT-1 for all loads $\rho$,
and prove that the gap asymptotically grows slower than $E[T^{SRPT \hy 1}]$.
By doing so, we will show that ServerFilling-SRPT is asymptotically optimal
in the multiserver-job system.

First, we prove a bound on the gap in mean response time between ServerFilling-SRPT
and SRPT-1:
\begin{theorem}
    \label{thm:sfs-bound}
    For all loads $\rho$,
    in the power-of-two setting,
    the mean response time gap between ServerFilling-SRPT
    and SRPT-1 is at most
    \begin{align*}
        E[T^{SFS \hy k}] - E[T^{SRPT \hy 1}] \le \frac{(e+1)(k-1)}{\lambda} 
            \ln \frac{1}{1-\rho} 
            +
            \frac{e}{\lambda}.
    \end{align*}
    The same is true of DivisorFilling-SRPT in the divisible setting.
\end{theorem}
\begin{proof}[Proof deferred to \cref{sec:proof}]\end{proof}

We use this bound to prove that ServerFilling-SRPT
yields optimal mean response time in the heavy-traffic limit:

\begin{theorem}
    \label{thm:sfs-opt}
    If $E[S^2 (\log S)^+] < \infty$,
    then ServerFilling-SRPT is asymptotically optimal in the multiserver-job system:
    \begin{align*}
        \lim_{\rho \to 1} \frac{E[T^{SFS \hy k}]}{E[T^{SRPT \hy 1}]}
        =
        \lim_{\rho \to 1} \frac{E[T^{SFS \hy k}]}{E[T^{OPT \hy k}]}
        =1.
    \end{align*}
    The same is true of DivisorFilling-SRPT in the divisible setting.
\end{theorem}
\begin{proof}[Proof deferred to \cref{sec:proof}]\end{proof}
The condition $E[S^2 (\log S)^+] < \infty$ is very slightly stronger than finite variance.

In \cref{sec:gittins-outline}, we generalize
both results to the settings of unknown- and partially-known job duration.
\subsection{A Novel Proof Technique: MIAOW}
\label{sec:miaow}
\subsubsection{Challenges of multiserver-job analysis}
As mentioned in \cref{sec:introduction},
mean response time analysis in the multiserver-job system
is a difficult problem,
with no size- or age-based scheduling policies having previously been analyzed.
The difficulty arises from two sources:
First, analyzing the mean response time of any
system with multiple servers
under a size- or age-based scheduling policy
is already very difficult,
even in a single-server-job setting such as the M/G/$k$.
New techniques based on relevant work
have recently been developed to handle this challenge.
The first such analysis is as recent as 2018,
when the SRPT-$k$ policy was analyzed in the M/G/$k$ \cite{grosof_srpt_2018},
followed by the analysis of the monotonic-Gittins-$k$ and Gittins-$k$ policies in the M/G/$k$
in 2020 and 2021 \cite{scully_optimal_2021,scully_gittins_2020}.

Unfortunately, the multiserver-job system presents a major additional challenge.
We will show in \cref{sec:first-attempt,sec:second-attempt}
that these recent techniques for multiserver systems
break when dealing with our multiserver-job system.
As a result, we need a new technique
to analyze the multiserver-job systems,
which we introduce in \cref{sec:our-approach}.

\subsubsection{Key idea of previous approaches: Relevant work similarity}
The first step in applying relevant-work-based techniques
\cite{grosof_srpt_2018,scully_optimal_2021,scully_gittins_2020}
is to prove a property which we call ``relevant work similarity'':
\begin{definition}
    A policy $\pi$ achieves \emph{relevant work similarity (RWS)}
    if, for all remaining sizes $r$ (or ranks%
\footnote{Rank is the analogue of remaining size under the Gittins policy.}
$r$),
    the policy $\pi$ system and the optimal resource-pooled system OPT-1
    (e.g. SRPT-1 or Gittins-1)
    have similar expected $r$-relevant work:
    \begin{align*}
        E[W^\pi_r] - E[W^{OPT \hy 1}_r] \le O(r).
    \end{align*}
\end{definition}
The RWS property holds for all three policies and systems analyzed previously
\cite{grosof_srpt_2018,scully_gittins_2020,scully_optimal_2021},
as well as for ServerFilling-SRPT.
Unfortunately, the RWS property is not sufficient
on its own to tightly bound mean response time,
or to prove asymptotically optimal mean response time.

\subsubsection{First attempt: Tagged job approach}
\label{sec:first-attempt}
One way to build on the RWS property to prove asymptotic optimality
is to use the tagged job approach,
employed by the SRPT-$k$ \cite{grosof_srpt_2018} and monotonic-Gittins-$k$
\cite{scully_optimal_2021} results.
The tagged job approach combines the RWS property with an additional property,
which we call ``relevant work implies response time'':
\begin{definition}
    A policy $\pi$ achieves \emph{relevant work implies response time (RW$\to$RT)} if
    the following holds: If a generic tagged job of size $r$ sees
    some amount $x$ of $r$-relevant work in each of the policy $\pi$ system
    and the optimal resource-pooled system OPT-1,
    then its expected response must be similar (within $O(r)$)
    in the two systems.
\end{definition}
If the RWS and RW$\to$RT properties can both be proven for some policy $\pi$,
it is relatively straightforward to tightly bound mean response time
and prove that the policy $\pi$ has asymptotically optimal mean response time.
Unfortunately, for our ServerFilling-SRPT policy,
the RW$\to$RT property fails,
meaning that the tagged-job approach cannot be used.

For a counterexample to the RW$\to$RT property for ServerFilling-SRPT,
consider a scenario where the tagged job
requires 1 server and has the smallest size of any job in the system,
and where it sees many jobs on arrival,
all of which require an even number of servers
and have larger remaining sizes.
Furthermore, assume that
arriving jobs rarely require 1 server.
The resource-pooled SRPT-1 system will quickly complete the tagged job,
as it has the smallest remaining size of any job in the system.

In contrast, the ServerFilling-SRPT system will not quickly complete the tagged job,
because ServerFilling-SRPT prioritizes the jobs of largest server need among the initial subset $M$,
as defined in \cref{sec:server-filling-defn}.
The tagged job will need to wait until the system empties or additional 1-server jobs arrive
to be served.
Clearly, similar relevant work does not imply similar response time.

This is an inherent difficulty of the multiserver-job system:
Serving the tagged job any earlier would require leaving at least one server empty,
as the tagged job is the only job with an odd server need,
given the power-of-two setting.
This could endanger throughput-optimality.
As a result, the tagged-job approach cannot be used
to effectively analyze the multiserver-job system.

\subsubsection{Second Attempt: Gittins-$k$}
\label{sec:second-attempt}

The analysis of the Gittins-$k$ policy for the M/G/$k$ \cite{scully_gittins_2020}
also relies on the RWS property,
which again is insufficient alone
to prove asymptotically optimal mean response time in their setting.
As in our setting, for the Gittins-$k$ system,
the RW$\to$RT property fails,
so the tagged-job approach cannot be employed.

The authors take a different approach:
They introduce WINE \cite[Theorem~6.3]{scully_gittins_2020},
our \cref{lem:wine},
a new identity that relates response time and relevant work
in all systems.%
\footnote{The name ``WINE'', short for ``work integral number equality'' \cite{scully_wine_2021},
is more recent than \cite{scully_gittins_2020}, but refers to their Theorem~6.3.}
WINE implies
\begin{align}
    \label{eq:wine}
    E[T^{\pi \hy k}] - E[T^{OPT \hy 1}] =
    \frac{1}{\lambda} \int_0^\infty \frac{E[W^{\pi \hy k}_r] - E[W^{OPT \hy 1}_r]}{r^2}.
\end{align}
WINE is more general than the RW$\to$RT property, because RW$\to$RT only holds in certain systems.

We can see from \eqref{eq:wine} that the RWS property is almost enough to bound mean response time,
but the $O(r)$ bound is too loose to show that the integral converges.
The authors therefore prove
a stronger version of the RWS property at sufficiently low and high ranks $r$.
Combining their strengthened bounds with WINE,
they prove that Gittins-$k$ achieves asymptotically optimal mean response time in the M/G/$k$.

However, their proof of a stronger version of RWS
at low ranks $r$ relies on the fact that under Gittins-$k$ in the M/G/$k$,
the job of least rank is guaranteed to be served.
This fails when applied to ServerFilling-SRPT,
because in our multiserver-job system
the job of least rank is not guaranteed to receive service.
See the counterexample given in \cref{sec:first-attempt}.

\subsubsection{Our approach}
\label{sec:our-approach}

Our key idea is to directly focus on the integrated relevant work difference
given in \eqref{eq:wine}.
This circumvents the need to strengthen the RWS property (like in \cref{sec:second-attempt})
or prove an RW$\to$RT property (like in \cref{sec:first-attempt}).

We start with a key property of the ServerFilling-SRPT system,
which we call ``relevant work efficiency'' (RWE).
RWE states that if there are $k$ or more $r$-relevant jobs
in the system,
then all servers are occupied by $r$-relevant jobs.
We prove in \cref{sec:server-filling-defn},
specifically in \cref{cor:key-sfs},
that ServerFilling-SRPT satisfies the RWE property.

While one can show that RWE implies RWS,
RWS alone is not enough,
as discussed in \cref{sec:second-attempt}.
Instead, we use the RWE property to directly bound the integrated relevant work difference
given in \eqref{eq:wine},
thereby directly bounding the mean response time difference.
We prove this result in \cref{thm:bound-idle}.
This forms the core of our proof that 
ServerFilling-SRPT achieves asymptotically optimal mean response time.

\cref{thm:bound-idle} is our key technical theorem;
it provides a novel bound on the \emph{waste}
in any system which satisfies RWE.
By waste, we refer to the quantity $E[W_r (1-B_r)]$,
the expected product of 
$r$-relevant work
and the fraction of system capacity not working on $r$-relevant jobs.
Note that the SRPT-1 system never has any waste:
If any $r$-relevant job is present
the entire system capacity is working on such a job.

To bound waste, we use a novel technique which we call MIAOW:
Multiplicative Interval Analysis of Waste.
Intuitively, MIAOW makes use of the fact that both $W_r$ and $B_r$
change slowly as a function of $r$.
We use this fact to bound
the integrated waste over a generic interval of remaining sizes $[r_\ell, r_h]$.
This contrasts with the prior waste-based technique \cite{scully_gittins_2020},
which focused on bounding waste at individual remaining sizes $r$,
an approach which does not imply a useful bound in the MSJ setting.
We then carefully select a sequence of remaining size intervals
with multiplicatively diminishing spare capacity $1-\rho^A_r$.
Applying our bound to each interval completes \cref{thm:bound-idle}.

We note that MIAOW
is stronger than the techniques used to prove asymptotically optimality in the M/G/k
for SRPT-$k$ and Gittins-$k$
\cite{grosof_srpt_2018,scully_gittins_2020}.
In particular, one could use our technique
to reprove all of the asymptotic optimality results in those papers.
This follows from the fact that the multiserver-job model is a generalization of the M/G/$k$:
A multiserver-job setting where all server needs are 1 is simply an M/G/$k$.
\subsection{Proof of Main Results}
\label{sec:proof}

Our goal is to bound the mean response time of the ServerFilling-SRPT policy,
relative to the resource-pooled SRPT-1 policy.

To bound mean response time,
we start by applying the ``work integral number equality'' (WINE)
technique \cite{scully_gittins_2020,scully_wine_2021}
to write mean response time $E[T^\pi]$ for a general policy $\pi$
in terms of expected relevant work $E[W_r^\pi]$.
This technique was introduced in \cite[Theorem~6.3]{scully_gittins_2020},
but we reprove it here for completeness.
\begin{lemma}[WINE Identity \cite{scully_gittins_2020}]
    \label{lem:wine}
    For an arbitrary scheduling policy $\pi$, in an arbitrary system,
    \begin{align*}
        E[T^\pi] = \frac{1}{\lambda} E[N^\pi] =
        \frac{1}{\lambda} \int_{r=0}^\infty \frac{E[W^\pi_r]}{r^2} dr.
    \end{align*}
\end{lemma}
\begin{proof}
    We will prove that at every moment in time,
    \begin{align}
        \label{eq:counting}
        N^\pi(t) = \int_{r=0}^\infty \frac{W^\pi_r (t)}{r^2} dr.
    \end{align}
    Recall that $r$-relevant work $W^\pi_r(t)$
    is simply a sum over the $r$-relevant jobs in the system.
    As a result, we can consider the integral in \eqref{eq:counting}
    as a sum over the jobs in the system.

    Consider a general job $j$, with remaining size $r_j$.
    The contribution of $j$ to the $r$-relevant work $W^\pi_r(t)$ is $r_j$,
    for thresholds $r$ such that $r_j \le r$,
    and 0 otherwise.

    Therefore, the contribution of job $j$ to the integral in \eqref{eq:counting}
    is
    \begin{align*}
        \int_{r=0}^\infty \frac{r_j \mathds{1}\{r_j \le r\}}{r^2} dr
        = \int_{r=r_j}^\infty \frac{r_j}{r^2} dr
        = r_j \int_{r=r_j}^\infty \frac{1}{r^2} dr
        = r_j \frac{1}{r_j} = 1.
    \end{align*}
    Because the contribution of an arbitrary job is 1,
    the integral in \eqref{eq:counting}
    simply counts the number of jobs in the system at time $t$,
    giving $N^\pi(t)$ as desired.

    Note that $E[T^\pi] = \frac{1}{\lambda} E[N^\pi]$,
    by Little's Law \cite{harchol_performance_2013}.
\end{proof}

Now that we have written mean response time in terms of relevant work,
we need to understand $E[W^\pi_r] - E[W^{SRPT \hy 1}_r]$,
the difference in $r$-relevant work between a general policy $\pi$
and the resource pooled SRPT-1 system.
To do so, we employ the work-decomposition law.
This technique was introduced in \cite{scully_gittins_2020},
and we specialize it here to the SRPT setting.
For completeness, we give the proof in \cref{app:work-decomp}.

\begin{lemma}{\cite[Theorem~7.2]{scully_gittins_2020}}
    \label{lem:decomp}
    For an arbitrary scheduling policy $\pi$,
    in an arbitrary known-size system,
    \begin{align*}
        E[W^\pi_r] - E[W^{SRPT \hy 1}_r]
        = \frac{E[(1-B^\pi_r) W^\pi_r] + \rho_r^R E_r[W_r^\pi]}{1-\rho_r^A}.
    \end{align*}
\end{lemma}
\begin{proof}[Proved in \cref{app:work-decomp}]
\end{proof}

Combining \cref{lem:wine}, and specifically its implication \eqref{eq:wine}, with \cref{lem:decomp},
we arrive at the following characterization of
the mean response time difference between a general policy $\pi$
and SRPT-1:
\begin{lemma}
\label{lem:mean-character}
    For any scheduling policy $\pi$, in any system,
    \begin{align}
    \label{eq:term-idle}
        E[T^\pi] - E[T^{SRPT \hy 1}]
        &= \frac{1}{\lambda}\int_0^\infty \frac{E[(1-B_r^{\pi})W_r^{\pi}]}{r^2(1-\rho^A_{r})} dr \\
    \label{eq:term-recycle}
        &+ \frac{1}{\lambda}\int_0^\infty \frac{\rho^R_r E_r[W_r^{\pi}]}{r^2(1-\rho^A_r)} dr.
    \end{align}
\end{lemma}

Intuitively, \eqref{eq:term-idle} and \eqref{eq:term-recycle}
measure the inefficiency of the policy $\pi$ relative to the ideal SRPT-1 system,
through the lens of $W^\pi_r$, the $r$-relevant work under policy $\pi$.

The first term \eqref{eq:term-idle} measures the extent to which
$r$-relevant work is present, but not being worked on.
In the multiserver-job system, not all of the system can be devoted to a single job,
so the waste $E[(1-B_r^{\pi})W_r^{\pi}]$ will typically be nonzero.

The second term \eqref{eq:term-recycle} measures the extent to which 
jobs $r$-recycle while $r$-relevant work is present in the system.
In the multiserver-job system, not all of the system can be devoted to a single job,
so jobs with remaining size above $r$ will be worked on,
and will $r$-recycle, while $r$-relevant work is present,
so $E_r[W_r^\pi]$ will also typically be nonzero.

Our goal is to bound the magnitude of \eqref{eq:term-idle} and \eqref{eq:term-recycle}
under the ServerFilling-SRPT policy, in the power-of-two setting.
We do so by making use of the key property of ServerFilling-SRPT,
relevant work efficiency (\cref{cor:key-sfs}):
If there are $k$ or more $r$-relevant jobs in the system, then $B_r = 1$.

We bound \eqref{eq:term-idle} in \cref{thm:bound-idle} using our novel MIAOW technique,
and we bound \eqref{eq:term-recycle} in \cref{thm:bound-recycle}.

\begin{theorem}[Bound waste]
    \label{thm:bound-idle}
    Under the ServerFilling-SRPT policy, in the power-of-two setting,
    \begin{align*}
        \int_{r=0}^\infty \frac{E[(1-B_r) W_r]}{r^2(1-\rho^A_r)} dr
        \le e(k-1) \biggl \lceil \ln \frac{1}{1-\rho} \biggr \rceil.
    \end{align*}
    The same is true of DivisorFilling-SRPT in the divisible setting.
\end{theorem}
\begin{proof}
    First, we make use of the key fact about ServerFilling-SRPT (and DivisorFilling-SRPT),
    relevant work efficiency:
    If there are at least $k$ jobs with rank $\le r$ in the system,
    then $B_r = 1$.
    This is proven in \cref{cor:key-sfs} for ServerFilling-SRPT,
    and in \cref{app:divisor-filling} for DivisorFilling-SRPT.

    Let us define $W^*_r$ to be the $r$-relevant work of the $k-1$ jobs of least remaining size in the system.
    Note that if $B_r < 1$, then $W_r = W^*_r$, for ServerFilling-SRPT and DivisorFilling-SRPT.
    As a result,
    \begin{align*}
        \int_{r=0}^\infty \frac{E[(1-B_r) W_r]}{r^2(1-\rho^A_r)} dr =
        \int_{r=0}^\infty \frac{E[(1-B_r) W^*_r]}{r^2(1-\rho^A_r)} dr.
    \end{align*}

    Next, we will break up the range of remaining sizes
    $r \in [0, \infty)$ into a finite set of buckets.
    Let $\{r_0, r_1, \ldots, r_m\}$ be a list of $m$ different remaining sizes, where $r_0 = 0$.
    We will specify the list $\{r_i\}$ later.
    Implicitly, we will say that $r_{m+1} = \infty$.
    We can rewrite the above integral as:
    \begin{align}
        \label{eq:bucketed}
        \int_{r=0}^\infty \frac{E[(1-B_r) W^*_r]}{r^2(1-\rho^A_r)} dr =
        \sum_{i=0}^m \int_{r=r_i}^{r_{i+1}} \frac{E[(1-B_r) W^*_r]}{r^2(1-\rho^A_r)} dr.
    \end{align}
    Next, we replace $r$ with either $r_i$ or $r_{i+1}$, selectively, to simplify things.
    Note that $B_r$ is increasing as a function of $r$, because as we increase the rank $r$,
    more servers are busy with $r$-relevant jobs.
    Likewise, $\rho^A_r$ is increasing as a function of $r$.
    Thus, for any $r \in [r_i, r_{i+1}]$,
    \begin{align*}
        B_{r_i} \le B_r, \qquad
        \rho^A_r \le \rho^A_{r_{i+1}}.
    \end{align*}
    Substituting into the integral from \eqref{eq:bucketed},
    we find that
    \begin{align*}
        \int_{r=r_i}^{r_{i+1}} \frac{E[(1-B_r) W^*_r]}{r^2(1-\rho^A_r)} dr
        \le \int_{r=r_i}^{r_{i+1}} \frac{E[(1-B_{r_i}) W^*_r]}{r^2(1-\rho^A_{r_{i+1}})} dr.
    \end{align*}
    Next, let us perform some algebraic manipulation:
    \begin{align}
        \label{eq:big-exp}
        \int_{r=r_i}^{r_{i+1}} \frac{E[(1-B_{r_i})W^*_r]}{r^2(1-\rho^A_{r_{i+1}})} dr
        =E \left[ \int_{r=r_i}^{r_{i+1}}
            \frac{(1-B_{r_i})W^*_r}{r^2(1-\rho^A_{r_{i+1}})} dr \right]
        =E \left[ \frac{1-B_{r_i}}{1-\rho^A_{r_{i+1}}} 
        \int_{r=r_i}^{r_{i+1}} \frac{W^*_r}{r^2} dr \right].
    \end{align}
    Now, let us make use of the definition of $W^*_r$.
    Recall that $W^*_r$ is the total remaining size
    of the $k-1$ jobs of least remaining size in the system.
    \begin{align*}
        W^*_r &= \sum_{j=1}^{k-1} r_j \mathds{1}\{r_j \le r\}
    \end{align*}
    Substituting this into \eqref{eq:big-exp},
    we find it is equal to
    \begin{align}
        \label{eq:service}
        = E \left[ \frac{1-B_{r_i}}{1-\rho^A_{r_{i+1}}}
            \sum_{j=1}^{k-1}
            \int_{r=r_i}^{r_{i+1}} \frac{r_j\mathds{1}\{r_j \le r\}}{r^2} dr \right].
    \end{align}
    Now, we will bound the integral in \eqref{eq:service}. As noted in \cref{lem:wine},
    for an arbitrary remaining size $r_j$,
    \begin{align*}
        \int_{r=0}^\infty \frac{r_j \mathds{1}\{r_j \le r\}}{r^2} dr
        = r_j \int_{r=r_j}^\infty \frac{1}{r^2} dr
        = r_j \frac{1}{r_j} = 1.
    \end{align*}
    As a result,
    \begin{align*}
        \int_{r=r_i}^{r_{i+1}} \frac{r_j\mathds{1}\{r_j \le r\}}{r^2} dr \le 1.
    \end{align*}

    Substituting in this bound into \eqref{eq:service},
    we find that
    \begin{align*}
        &E \left[ \frac{1-B_{r_i}}{1-\rho^A_{r_{i+1}}}
        \sum_{j=1}^{k-1} \int_{r=r_i}^{r_{i+1}}
        \frac{r_j\mathds{1}\{r_j \le r\}}{r^2} dr \right]
        \le E \left[ \frac{1-B_{r_i}}{1-\rho^A_{r_{i+1}}} (k-1) \right] \\
        &= (k-1) E \left[ \frac{1-B_{r_i}}{1-\rho^A_{r_{i+1}}} \right]
        = (k-1) \frac{1-\rho_{r_i}}{1-\rho^A_{r_{i+1}}}
        \le (k-1) \frac{1-\rho^A_{r_i}}{1-\rho^A_{r_{i+1}}}.
    \end{align*}
    Returning all the way back to the beginning, we find that
    \begin{align}
    \label{eq:final}
        \int_{r=0}^\infty \frac{E[(1-B_r) W_r]}{r^2(1-\rho^A_r)} dr
        \le (k-1)\sum_{i=0}^m \frac{1-\rho^A_{r_i}}{1-\rho^A_{r_{i+1}}}.
    \end{align}
    We are now ready to construct the list $\{r_i\}$.
    Our goal in doing so is to minimize the sum
    \begin{align*}
        \sum_{i=0}^m \frac{1-\rho^A_{r_i}}{1-\rho^A_{r_{i+1}}}.
    \end{align*}
    Our only constrains are that $r_0 = 0$ and $r_{m+1}=\infty$.
    In particular,
    \begin{align*}
        1-\rho^A_{r_0} = 1 - \rho^A_0 = 1, \qquad
        1 - \rho^A_{r_{m+1}} = 1 - \rho^A_\infty = 1 - \rho.
    \end{align*}
    All other $r_i$ thresholds are ours to choose.

    We will set $r_i$ such that the values
    $1 - \rho^A_{r_i}$ form a geometric progression.
    In particular,
    define $r_1, r_2, \ldots$ to satisfy the following:
    \begin{align}
        \label{eq:choose-r}
        1-\rho_{r_1}^A = \frac{1}{e}, \qquad
        1-\rho_{r_2}^A = \frac{1}{e^2}, \qquad
        \ldots \qquad
        1-\rho_{r_i}^A = \frac{1}{e^i} \quad \forall i \le m.
    \end{align}
    If the size distribution $S$ is continuous,
    we choose $r_i$ to exactly satisfy \eqref{eq:choose-r}. 
    If $S$ is discontinuous,
    then $\rho_r^A$ is discontinuous, so
    exact equality is not necessarily possible. However, it suffices to choose $r_i$ such that
    \begin{align*}
        \frac{1}{e^i} \in [1 - \rho_{r_i^+}^A, 1-\rho_{r_i^-}^A] \quad \forall i \le m,
    \end{align*}
    which is always possible.
    By $^+$ and $^-$, we refer to the one-sided limits.

    We then set $m = \lceil \ln \frac{1}{1-\rho} \rceil - 1$.
    This choice of $\{r_i\}$ ensures that
    \begin{align}
    \label{eq:ratios}
        \frac{1-\rho^A_{r_i}}{1-\rho^A_{r_{i+1}}} &\le e \quad \forall i \le m \\
        \sum_{i=0}^m \frac{1-\rho^A_{r_i}}{1-\rho^A_{r_{i+1}}} &\le e(m+1) = e\biggl \lceil \ln \frac{1}{1-\rho} \biggr \rceil.
    \end{align}
    For $i \le m-1$, \eqref{eq:ratios} follows immediately from \eqref{eq:choose-r}.
    For $i=m$, \eqref{eq:ratios} follows from the fact that $1-\rho^A_{r_{m+1}} = 1-\rho$.

    Applying \eqref{eq:final}, we find that
    \begin{align*}
        \int_{r=0}^\infty \frac{E[(1-B_r) W_r]}{r^2(1-\rho^A_{r})} dr
        \le e(k-1)\biggl \lceil \ln \frac{1}{1-\rho} \biggr \rceil.
        \tag*{\qedhere}
    \end{align*}
\end{proof}
Now, it remains to bound \eqref{eq:term-recycle}:
\begin{theorem}[Bound recycled work]
    \label{thm:bound-recycle}
    Under the ServerFilling-SRPT policy, in the power-of-two setting,
    \begin{align*}
        \int_{r=0}^\infty \frac{\rho^R_r E_r[W_r]}{r^2(1-\rho^A_{r})} dr
        \le (k-1) \ln \frac{1}{1-\rho}.
    \end{align*}
    The same is true of DivisorFilling-SRPT in the divisible setting.
\end{theorem}
\begin{proof}
    First, recall the key property of ServerFilling-SRPT and DivisorFilling-SRPT,
    relevant work efficiency:
    If there are at least $k$ jobs with remaining size $\le r$ in the system, then $B_r = 1$.
    This is proven in \cref{cor:key-sfs} for ServerFilling-SRPT,
    and in \cref{app:divisor-filling} for DivisorFilling-SRPT.

    When a job $r$-recycles,
    it must have been in service despite having remaining size $> r$.
    As a result, there are at most $k-1$ other jobs with remaining size $\le r$
    present in the system at an $r$-recycling moment.
    Each such job contributes at most $r$ work to $W_r$.
    As a result,
    $E_r[W_r] \le (k-1) r$.
    \begin{align*}
        \int_{r=0}^\infty \frac{\rho^R_r E_r[W_r]}{r^2(1-\rho^A_{r})} dr
        \le \int_{r=0}^\infty \frac{(k-1)r \rho^R_r}{r^2(1-\rho^A_{r})} dr
        = (k-1) \int_{r=0}^\infty \frac{\rho^R_r}{1-\rho^A_{r}} \frac{1}{r} dr.
    \end{align*}
    To bound the integrand, we will expand the definitions of $\rho^R_r$
    and $\rho^A_{r}$ in the SRPT setting.
    \begin{align*}
        \rho^R_r &= \lambda r P(S > r) \\
        \rho^A_{r} &= \lambda E[S \mathds{1}\{S \le r\}].
    \end{align*}

    We therefore bound as follows:
    \begin{align*}
        \frac{\rho^R_r}{1-\rho^A_{r}}
        \frac{1}{r}
        =
        \frac{\lambda r P(S > r)}{1 - \lambda E[S \mathds{1}\{S \le r\}]}
        \frac{1}{r}
        = \frac{\lambda P(S > r)}{1 - \lambda E[S \mathds{1}\{S \le r\}]}.
    \end{align*}
    Now, note that $P(S > r) = \frac{d}{dr} E[\min(S, r)]$,
    and that $E[\min(S, r)] \ge E[S \mathds{1}\{S \le r\}]$.
    As a result,
    \begin{align*}
        \frac{\lambda P(S > r)}{1 - \lambda E[S \mathds{1}\{S \le r\}]}
        \le \frac{\lambda P(S > r)}{1 - \lambda E[\min(S, r)]}
        = \frac{\lambda \frac{d}{dr} E[\min(S, r)]}{1 - \lambda E[\min(S, r)]}
        = - \frac{d}{dr} \ln \frac{1}{1-\lambda E[\min(S, r)]}.
    \end{align*}

    Integrating over all $r \in [0, \infty)$, we find that
    \begin{align*}
        \int_{r=0}^\infty \frac{\rho^R_r}{1-\rho^A_{r})} \frac{1}{r} dr
        \le \left[ - \ln \frac{1}{1 - \lambda E[\min(S, r)]} \right]_{r=0}^\infty
        = \ln \frac{1}{1-\rho}.
        \tag*{\qedhere}
    \end{align*}
\end{proof}

Now, we're ready to put it all together.
We derive a bound on mean response time:
\begin{reptheorem}{thm:sfs-bound}
    In any multiserver-job system,
    the difference in mean response time between ServerFilling-SRPT
    and SRPT-1 is at most
    \begin{align*}
        E[T^{SFS \hy k}] - E[T^{SRPT \hy 1}]
            \le  \frac{(e+1)(k-1)}{\lambda} \ln \left( \frac{1}{1-\rho} \right) + \frac{e}{\lambda}.
    \end{align*}
    The same is true of DivisorFilling-SRPT in the divisible setting.
\end{reptheorem}
\begin{proof}
    From \cref{lem:mean-character}, we know that
    \begin{align*}
        &E[T^{SFS \hy k}] - E[T^{SRPT \hy 1}] \\
        &= \frac{1}{\lambda}
        \int_0^\infty \frac{E[(1-B_r^{SFS \hy k})W_r^{SFS \hy k}]}{r^2(1-\rho^A_{r})} dr
        + \frac{1}{\lambda}
        \int_0^\infty \frac{\rho^R_rE_r[W_r^{SFS \hy k}]}{r^2(1-\rho^A_{r})} dr.
    \end{align*}
    We apply \cref{thm:bound-idle} and \cref{thm:bound-recycle} to bound the two terms:
    \begin{align*}
        &E[T^{SFS \hy k}] - E[T^{SRPT \hy 1}]
        \le \frac{1}{\lambda}e(k-1)\biggl \lceil \ln \frac{1}{1-\rho} \biggr \rceil
        + \frac{1}{\lambda}(k-1) \ln \frac{1}{1-\rho}.
    \end{align*}
    We use the bound $\lceil x \rceil \le x + 1$
    to simplify the resulting expression.
\end{proof}

Now, we use this bound to prove asymptotic optimality:
\begin{reptheorem}{thm:sfs-opt}
    If $E[S^2(\log S)^+] < \infty$,
    \begin{align*}
        \lim_{\rho \to 1} \frac{E[T^{SFS \hy k}]}{E[T^{SRPT \hy 1}]} = 
        \lim_{\rho \to 1} \frac{E[T^{SFS \hy k}]}{E[T^{OPT \hy k}]} = 
        1.
    \end{align*}
    The same is true of DivisorFilling-SRPT in the divisible setting.
\end{reptheorem}
\begin{proof}
From \cref{thm:sfs-bound},
we know that the gap $E[T^{SFS \hy k}] - E[T^{SRPT \hy 1}]$
grows as $O(\log \frac{1}{1-\rho})$ in the $\rho \to 1$ limit.
It is known that
if $E[S^2(\log S)^+] < \infty$,
then $E[T^{SRPT \hy 1}] = \omega(\log \frac{1}{1-\rho})$
in the $\rho \to 1$ limit.
This is proven in \cite[Appendix~B.2]{scully_gittins_2020},
and specifically in the proof of \cite[Theorem~1.3]{scully_gittins_2020}.
\end{proof}
\section{ServerFilling-Gittins: Asymptotic Optimality with Unknown Sizes}
\label{sec:gittins-outline}
We generalize our results to the setting of unknown sizes or of partially known sizes
(e.g. size estimates).
To do so, we replace the SRPT job ordering with the Gittins job ordering,
thus creating the ServerFilling-Gittins (SFG-$k$) and DivisorFilling-Gittins policies.
\subsection{Background}
The Gittins policy is the optimal scheduling policy
for minimizing mean response time in the M/G/1 in the unknown and partially-known size settings
\cite{gittins_multi_2011,scully_gittins_2021},
filling the same role as SRPT in the known-size setting.

The Gittins policy is an age-based index policy, meaning that it assigns
each job a rank according to the job's age and static characteristics (e.g. server need),
as well as any other information the scheduler may have,
and serves the job of least rank.
In the blind MSJ setting, the Gittins rank function can be defined as follows:
Let $S_i$ be the job size distribution of jobs with server need $i$.
Then a job with server need $i$ and age $a$ has rank:
\begin{align*}
    \inf_{b > a} \frac{E[\min(S_i, b) - a \mid S_i > a]}{P[S_i \le b | S_i > a]}.
\end{align*}
The definition of the Gittins rank in settings where the server has more information
is similar, but more complicated. For more details, see \cite{scully_gittins_2020,scully_gittins_2021}.

We define the ServerFilling-Gittins policy by ordering jobs in increasing order of Gittins rank,
and then applying the same ServerFilling procedure as described in \cref{sec:server-filling-defn}.
We define DivisorFilling-Gittins similarly,
based on the DivisorFilling procedure given in \cref{app:divisor-filling}.

\subsection{Notation}
\label{sec:gittins-notation}

Our notation follows \cite{scully_gittins_2020}.
We start by defining a job state space $X$ of all possible job states $x$.
For instance, in the unknown size setting, a job's state is simply its age $a$.
In the known-size setting, a job's state was its remaining size.
Every state $x$ is mapped to $\mathrm{rank}(x)$.
We call a job in state $x$ $r$-relevant if $\mathrm{rank}(x) < r$.

Next, we need to adjust the concept of ``remaining size'' slightly.
We define $S_r(x)$, the \emph{$r$-relevant remaining size} of a job in state $x$,
to be the random variable denoting the amount of service the job needs in order
to reach an $r$-irrelevant state or complete.
In the known-size case, this amount of service was deterministic,
but here it is a random variable.

We can now define $W_r$,
the \emph{$r$-relevant work} in the system,
to be the total of all jobs' $r$-relevant remaining size in steady state.
Likewise, $B_r$ is the fraction of servers occupied by $r$-relevant jobs.

We also define two state distributions:
$X^A$, the state of arriving jobs,
and $X^R_r$, the state of jobs recycling relative to rank $r$.
In the known-size case, $X^R_r$ is deterministic,
and in the unknown size case, $X^A$ is deterministic,
but in general both are random variables.
We also define $\lambda^R_r$ to be the rate at which jobs recycle relative to rank $r$.
This is equal to $\lambda$ times the expected number of $r$-recyclings per job.

We can now define the two constituents of $r$-relevant load,
$\rho_r^A$ and $\rho_r^R$.
\begin{align*}
    \rho^A_r &:= \lambda E[S_r(X^A)] \\
    \rho^R_r &:= \lambda^R_r E_r[S_r(X^R_r)]
\end{align*}

Now, we are ready to state our main result for ServerFilling-Gittins.

\subsection{Asymptotic Optimality for ServerFilling-Gittins}
Our main result for ServerFilling-Gittins is an analogous bound on mean response time to
\cref{thm:sfs-bound}, our bound on mean response time for ServerFilling-SRPT:
\begin{theorem}
    \label{thm:gittins-bound}
    For all loads $\rho$,
    in the power-of-two setting,
    the mean response time gap between ServerFilling-Gittins
    and Gittins-1 is at most
    \begin{align*}
        E[T^{SFG \hy k}] - E[T^{Gittins \hy 1}] \le \frac{(e+1)(k-1)}{\lambda} 
            \ln \frac{1}{1-\rho} 
            +
            \frac{e}{\lambda}.
    \end{align*}
    The same is true of DivisorFilling-Gittins in the divisible setting.
\end{theorem}
Note that this bound is in some ways stronger than the bound on Gittins-$k$
given in \cite{scully_gittins_2020}.
Our bound is the first uniform bound on multiserver Gittins,
meaning that our bound doesn't depend on $S$ except via $E[S]$,
unlike the bound on Gittins-$k$ in \cite{scully_gittins_2020}.
Note also that the M/G/$k$ is a special case of the multiserver-job system when server needs are all 1,
and that in this special case, ServerFilling-Gittins specializes to Gittins-$k$.
As a result, \cref{thm:gittins-bound} is a strict improvement
upon the bound given in \cite{scully_gittins_2020}.

We use this bound to prove that ServerFilling-Gittins
yields optimal mean response time in the heavy-traffic limit:

\begin{theorem}
    \label{thm:gittins-opt}
    If $E[S^2 (\log S)^+] < \infty$,
    then ServerFilling-Gittins is asymptotically optimal in the multiserver-job system:
    \begin{align*}
        \lim_{\rho \to 1} \frac{E[T^{SFG \hy k}]}{E[T^{Gittins \hy 1}]}
        =
        \lim_{\rho \to 1} \frac{E[T^{SFG \hy k}]}{E[T^{OPT \hy k}]}
        =1.
    \end{align*}
    The same is true of DivisorFilling-Gittins in the divisible setting.
\end{theorem}
\cref{thm:gittins-opt} follows from \cref{thm:gittins-bound}
just as \cref{thm:sfs-opt} follows from \cref{thm:sfs-bound}.

To prove \cref{thm:gittins-bound},
an analogous proof to the proof of \cref{thm:sfs-bound} given in \cref{sec:proof}
suffices.
We simply must replace certain quantities used in \cref{sec:proof}
with the equivalent quantities for the Gittins policy.
Specifically, rather than thinking of a job as $r$-relevant if it has remaining size $\le r$,
we instead think of a job as $r$-relevant if it has rank $\le r$
under the Gittins policy.
We redefine $W^\pi_r$, $B^\pi_r$, and $\rho_r$, $\rho_r^A$, and $\rho_r^R$ accordingly,
as described in \cref{sec:gittins-notation}.
For full details, see \cref{app:gittins}.

The recycling term of our key background lemma \cref{lem:gittins-mean-character} is likewise slightly different:
\begin{lemma}
\label{lem:gittins-mean-character}
    For any scheduling policy $\pi$,
    \begin{align}
        E[T^\pi] - E[T^{Gittins \hy 1}]
        &= \frac{1}{\lambda}\int_0^\infty \frac{E[(1-B_r^{\pi})W_r^{\pi}]}{r^2(1-\rho^A_{r})} dr \\
        &+ \frac{1}{\lambda}\int_0^\infty \frac{\lambda^R_r E_r[S_r(X^R_r)W_r^{\pi}]}{r^2(1-\rho^A_r)} dr.
    \end{align}
\end{lemma}
Here $\rho_R^r E_r[W^\pi_r]$ from \cref{lem:mean-character}
becomes $\lambda^R_r E_r[S_r(X^R_r)W_r^{\pi}]$.
Note that in the SRPT case, $S_r(X^R)=r$,
because under SRPT, a job $r$-recycles when its remaining size is $r$.
\cref{lem:gittins-mean-character} follows from \cite[Theorem 7.2]{scully_gittins_2020}.

Bounding the waste term involving $E[(1-B_r^{\pi})W_r^{\pi}]$
proceeds completely analogously to \cref{thm:bound-idle}.
Bounding the recycled work term involving $\lambda^R_r E_r[S_r(X^R_r)W_r^{\pi}]$
is likewise completely analogous to \cref{thm:bound-recycle}.
For the full details, see \cref{app:gittins}.
\section{Empirical Results}
\label{sec:empirical}
We have proven that ServerFilling-SRPT yields asymptotically optimal mean response time in the heavy-traffic limit (as $\rho \to 1$).
To empirically validate our theoretical results and broaden our comparison to general $\rho$,
we use simulation to compare the mean response time of ServerFilling-SRPT
to that of several previously proposed policies:
\begin{description}
    \item[MaxWeight:] A throughput optimal policy 
    which considers all possible sets of jobs that can be served at a time.
    Each job is given a weight equal the number of jobs in the system
    with the same server need.
    The set of jobs with the maximum total weight is served \cite{maguluri_2012}.
    Note that this policy requires solving a NP-hard Bin Packing problem for each service.
    \item[ServerFilling:] A policy which orders jobs in arrival order,
    then uses the same procedure to place jobs onto servers as our ServerFilling-SRPT
    policy specified in \cref{sec:server-filling-defn}.
    ServerFilling is throughput-optimal in the power-of-two setting \cite{grosof_wcfs_2021}.
\end{description}

We also compare against
resource-pooled SRPT-1, our lower bound on the optimal policy.

\begin{figure}
    \centering
    \includegraphics[width=0.8\textwidth]{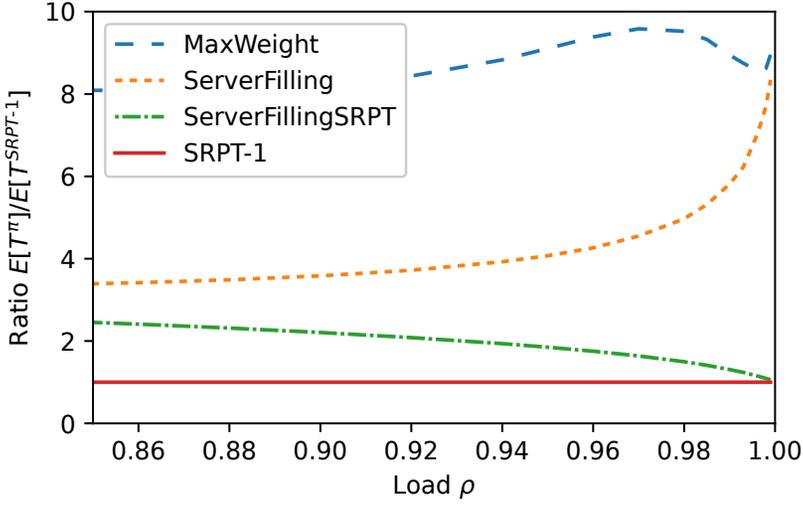}
    \caption{Ratio of mean response time between several multiserver-job policies and SRPT-1.
        $K$ uniformly sampled from $\{1, 2, 4, 8\}$.
        $S$ exponentially distributed, independent of $K$.
        Each simulation consists of $10^7$ arrivals.
        Loads up to $\rho = 0.999$ simulated.}
    \label{fig:ratio-exp}
\end{figure}

In \cref{fig:ratio-exp},
we show the ratio of mean response time between the multiserver-job policies and SRPT-1.
As proven in \cref{thm:sfs-opt}, for ServerFilling-SRPT, this ratio converges to 1,
implying that ServerFilling-SRPT yields asymptotically optimal mean response time.
In contrast, for MaxWeight and ServerFilling, the ratio is far from one, and appears to diverge.
ServerFilling-SRPT has superior mean response time at all $\rho$.

\begin{figure}
    \centering
    \includegraphics[width=0.8\textwidth]{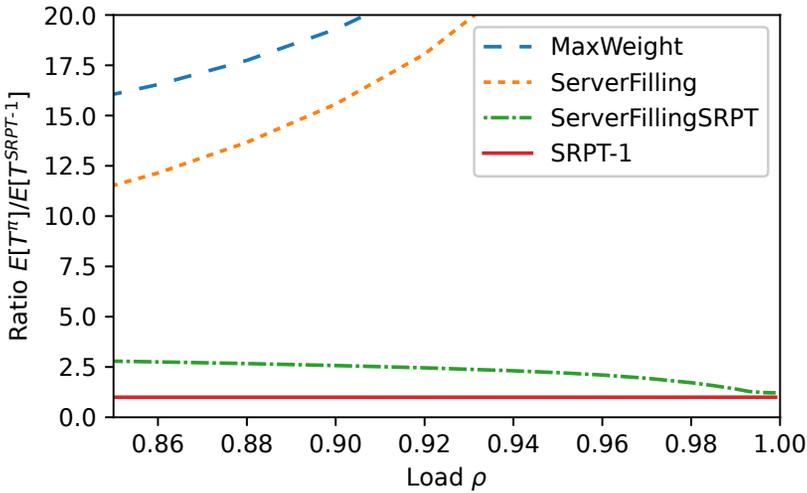}
    \caption{Ratio of mean response time between several multiserver-job policies and SRPT-1
    under high variance.
        $K$ uniformly sampled from $\{1, 2, 4, 8\}$.
        $S$ hyperexponentially distributed, $C^2=10$, independent of $K$.
        Each simulation consists of $10^7$ arrivals.
        Loads up to $\rho = 0.999$ simulated.}
    \label{fig:ratio-hyp}
\end{figure}

In \cref{fig:ratio-hyp},
we show a setting with higher variance job sizes, where $C^2 = 10$.
In high-variance settings, making effective use of job size information is at its most important.
Here, the ratio for ServerFilling-SRPT again converges smoothly to 1,
while the ratios for MaxWeight and ServerFilling diverge rapidly.

In \cref{sec:introduction},
\cref{fig:intro-mean},
we also compared ServerFilling-SRPT against two size-based heuristic policies:
\begin{description}
    \item[GreedySRPT:] 
        Order jobs in increasing order of remaining size.
        As long as sufficient servers are available,
        place jobs into service.
        When a job has higher server need than the remaining number of servers available,
        stop.
    \item[FirstFitSRPT:]
        Order jobs in increasing order of remaining size.
        As long as sufficient servers are available,
        place jobs into service.
        If a job has higher server need than the remaining number of servers available,
        skip that job. Continue through the list of jobs, placing jobs into service
        if sufficient servers are available, until all servers are full,
        or all jobs are exhausted.
        This policy was studied under the name ``Smallest Area First'' \cite{carastan_one_2019}.
\end{description}
GreedySRPT makes no effort to pack jobs efficiently onto servers,
while FirstFitSRPT is unreliable at doing so.
For both of these policies, the stability
region is significantly smaller than the optimal stability region.
This is why
neither policy is depicted in \cref{fig:ratio-exp} or \cref{fig:ratio-hyp},
as both are unstable for all loads $\rho \ge 0.85$,
and hence have infinite mean response time on this domain.

We summarize our experiments as follows:
In all experiments, at all $\rho$, ServerFilling-SRPT has minimal
mean response time.
\section{Conclusion}
We introduce the ServerFilling-SRPT scheduling policy for the multiserver-job system.
We prove a tight bound on the mean response time
of ServerFilling-SRPT in the power-of-two setting, which applies for all loads $\rho$.
We use that bound to prove that
ServerFilling-SRPT achieves asymptotically optimal mean response time in heavy traffic.
We also show that ServerFilling-SRPT empirically achieves the best mean response time of any policy simulated, across all loads $\rho$.
We also introduce the DivisorFilling-SRPT policy, in the more general divisible setting,
and the ServerFilling- and DivisorFilling-Gittins policies,
in the settings of unknown- and partially-known job sizes,
proving similar asymptotic optimality results for each.

One of the major insights of this paper is that achieving asymptotically optimal mean response time
requires prioritizing jobs of small remaining size
without sacrificing the throughput of the system.
ServerFilling-SRPT is the first policy to achieve both goals simultaneously.

The MIAOW analysis technique introduced in this paper
extends beyond ServerFilling-SRPT and the multiserver-job setting.
In fact, it allows the analysis of any system and any policy in which the relevant work efficiency
property (\cref{cor:key-sfs}) can be proven.

One direction of future work is to study multiserver-job scheduling policies
outside of the divisible setting.
No mean response time analysis is currently known for any scheduling policy
in this more general setting,
much less any optimality results,
so new techniques will likely be needed.
In particular, no policy with the remaining work efficiency property can exist in this setting.

\section{Acknowledgements}

This research was supported by NSF-CMMI-1938909 and NSF-CSR-1763701.
Isaac Grosof was supported by a Siebel Scholar Award 2022-2023.
Part of this work was done while Ziv Scully was visiting
the Simons Institute for the Theory of Computing,
where he was supported by a VMware research fellowship.
We thank the reviewers for valuable feedback.
We thank Katherine Kosaian for aid in creating the DivisorFilling,
DivisorFilling-SRPT, and DivisorFilling-Gittins policies.

\bibliographystyle{ACM-Reference-Format}
\bibliography{refs}


\begin{thebibliography}{43}


\ifx \showCODEN    \undefined \def \showCODEN     #1{\unskip}     \fi
\ifx \showDOI      \undefined \def \showDOI       #1{#1}\fi
\ifx \showISBNx    \undefined \def \showISBNx     #1{\unskip}     \fi
\ifx \showISBNxiii \undefined \def \showISBNxiii  #1{\unskip}     \fi
\ifx \showISSN     \undefined \def \showISSN      #1{\unskip}     \fi
\ifx \showLCCN     \undefined \def \showLCCN      #1{\unskip}     \fi
\ifx \shownote     \undefined \def \shownote      #1{#1}          \fi
\ifx \showarticletitle \undefined \def \showarticletitle #1{#1}   \fi
\ifx \showURL      \undefined \def \showURL       {\relax}        \fi
\providecommand\bibfield[2]{#2}
\providecommand\bibinfo[2]{#2}
\providecommand\natexlab[1]{#1}
\providecommand\showeprint[2][]{arXiv:#2}

\bibitem[{Armstrong} et~al\mbox{.}(2010)]%
        {armstrong_scheduling}
\bibfield{author}{\bibinfo{person}{Timothy~G. {Armstrong}},
  \bibinfo{person}{Zhao {Zhang}}, \bibinfo{person}{Daniel~S. {Katz}},
  \bibinfo{person}{Michael {Wilde}}, {and} \bibinfo{person}{Ian~T. {Foster}}.}
  \bibinfo{year}{2010}\natexlab{}.
\newblock \showarticletitle{Scheduling many-task workloads on supercomputers:
  Dealing with trailing tasks}. In \bibinfo{booktitle}{\emph{2010 3rd Workshop
  on Many-Task Computing on Grids and Supercomputers}}. \bibinfo{pages}{1--10}.
\newblock


\bibitem[Arthurs and Kaufman(1979)]%
        {arthurs_sizing_1979}
\bibfield{author}{\bibinfo{person}{Edward Arthurs} {and}
  \bibinfo{person}{Joseph~S. Kaufman}.} \bibinfo{year}{1979}\natexlab{}.
\newblock \showarticletitle{Sizing a message store subject to blocking
  criteria}. In \bibinfo{booktitle}{\emph{Proceedings of the third
  international symposium on modelling and performance evaluation of computer
  systems: Performance of computer systems}}. \bibinfo{pages}{547--564}.
\newblock


\bibitem[Brill and Green(1984)]%
        {brill_queues_1984}
\bibfield{author}{\bibinfo{person}{Percy~H. Brill} {and} \bibinfo{person}{Linda
  Green}.} \bibinfo{year}{1984}\natexlab{}.
\newblock \showarticletitle{Queues in Which Customers Receive Simultaneous
  Service from a Random Number of Servers: A System Point Approach}.
\newblock \bibinfo{journal}{\emph{Management Science}} \bibinfo{volume}{30},
  \bibinfo{number}{1} (\bibinfo{year}{1984}), \bibinfo{pages}{51--68}.
\newblock


\bibitem[Carastan-Santos et~al\mbox{.}(2019)]%
        {carastan_one_2019}
\bibfield{author}{\bibinfo{person}{Danilo Carastan-Santos},
  \bibinfo{person}{Raphael~Y. De~Camargo}, \bibinfo{person}{Denis Trystram},
  {and} \bibinfo{person}{Salah Zrigui}.} \bibinfo{year}{2019}\natexlab{}.
\newblock \showarticletitle{One Can Only Gain by Replacing {EASY} Backfilling:
  A Simple Scheduling Policies Case Study}. In \bibinfo{booktitle}{\emph{2019
  19th IEEE/ACM International Symposium on Cluster, Cloud and Grid Computing
  (CCGRID)}}. \bibinfo{pages}{1--10}.
\newblock


\bibitem[Cirne and Berman(2001)]%
        {cirne_model_2001}
\bibfield{author}{\bibinfo{person}{Walfredo Cirne} {and}
  \bibinfo{person}{Francine Berman}.} \bibinfo{year}{2001}\natexlab{}.
\newblock \showarticletitle{A model for moldable supercomputer jobs}. In
  \bibinfo{booktitle}{\emph{Proceedings 15th International Parallel and
  Distributed Processing Symposium. IPDPS 2001}}. \bibinfo{pages}{8 pp.}
\newblock


\bibitem[Downey(1997)]%
        {downey_using_1997}
\bibfield{author}{\bibinfo{person}{Allen~B. Downey}.}
  \bibinfo{year}{1997}\natexlab{}.
\newblock \showarticletitle{Using queue time predictions for processor
  allocation}. In \bibinfo{booktitle}{\emph{workshop on Job Scheduling
  Strategies for Parallel Processing}}. Springer, \bibinfo{pages}{35--57}.
\newblock


\bibitem[Etsion and Tsafrir(2005)]%
        {etsion_short}
\bibfield{author}{\bibinfo{person}{Yoav Etsion} {and} \bibinfo{person}{Dan
  Tsafrir}.} \bibinfo{year}{2005}\natexlab{}.
\newblock \showarticletitle{A short survey of commercial cluster batch
  schedulers}.
\newblock \bibinfo{journal}{\emph{School of Computer Science and Engineering,
  The Hebrew University of Jerusalem}}  \bibinfo{volume}{44221}
  (\bibinfo{year}{2005}), \bibinfo{pages}{2005--13}.
\newblock


\bibitem[Feitelson and Rudolph(1996)]%
        {feitelson_toward}
\bibfield{author}{\bibinfo{person}{Dror~G. Feitelson} {and}
  \bibinfo{person}{Larry Rudolph}.} \bibinfo{year}{1996}\natexlab{}.
\newblock \showarticletitle{Toward convergence in job schedulers for parallel
  supercomputers}. In \bibinfo{booktitle}{\emph{Job Scheduling Strategies for
  Parallel Processing}}, \bibfield{editor}{\bibinfo{person}{Dror~G. Feitelson}
  {and} \bibinfo{person}{Larry Rudolph}} (Eds.). \bibinfo{publisher}{Springer
  Berlin Heidelberg}, \bibinfo{address}{Berlin, Heidelberg},
  \bibinfo{pages}{1--26}.
\newblock
\showISBNx{978-3-540-70710-3}


\bibitem[Feitelson et~al\mbox{.}(2004)]%
        {feitelson_parallel}
\bibfield{author}{\bibinfo{person}{Dror~G. Feitelson}, \bibinfo{person}{Larry
  Rudolph}, {and} \bibinfo{person}{Uwe Schwiegelshohn}.}
  \bibinfo{year}{2004}\natexlab{}.
\newblock \showarticletitle{Parallel job scheduling--a status report}. In
  \bibinfo{booktitle}{\emph{Workshop on Job Scheduling Strategies for Parallel
  Processing}}. Springer, \bibinfo{pages}{1--16}.
\newblock


\bibitem[Filippopoulos and Karatza(2007)]%
        {fillippopoulos_mm2}
\bibfield{author}{\bibinfo{person}{Dimitrios Filippopoulos} {and}
  \bibinfo{person}{Helen Karatza}.} \bibinfo{year}{2007}\natexlab{}.
\newblock \showarticletitle{An M/M/2 parallel system model with pure space
  sharing among rigid jobs}.
\newblock \bibinfo{journal}{\emph{Mathematical and Computer Modelling}}
  \bibinfo{volume}{45}, \bibinfo{number}{5} (\bibinfo{year}{2007}),
  \bibinfo{pages}{491 -- 530}.
\newblock
\showISSN{0895-7177}


\bibitem[Ghaderi(2016)]%
        {ghaderi_randomized_2016}
\bibfield{author}{\bibinfo{person}{Javad Ghaderi}.}
  \bibinfo{year}{2016}\natexlab{}.
\newblock \showarticletitle{Randomized algorithms for scheduling {VMs} in the
  cloud}. In \bibinfo{booktitle}{\emph{{IEEE} {INFOCOM} 2016 - {The} 35th
  {Annual} {IEEE} {International} {Conference} on {Computer}
  {Communications}}}. \bibinfo{pages}{1--9}.
\newblock


\bibitem[Gittins et~al\mbox{.}(2011)]%
        {gittins_multi_2011}
\bibfield{author}{\bibinfo{person}{John Gittins}, \bibinfo{person}{Kevin
  Glazebrook}, {and} \bibinfo{person}{Richard Weber}.}
  \bibinfo{year}{2011}\natexlab{}.
\newblock \bibinfo{booktitle}{\emph{Multi-armed bandit allocation indices}}.
\newblock \bibinfo{publisher}{John Wiley \& Sons}.
\newblock


\bibitem[Grosof et~al\mbox{.}(2021)]%
        {grosof_wcfs_2021_full}
\bibfield{author}{\bibinfo{person}{Isaac Grosof}, \bibinfo{person}{Mor
  Harchol-Balter}, {and} \bibinfo{person}{Alan Scheller-Wolf}.}
  \bibinfo{year}{2021}\natexlab{}.
\newblock \showarticletitle{WCFS: A new framework for analyzing multiserver
  systems}.
\newblock \bibinfo{journal}{\emph{arXiv preprint arXiv:2109.12663}}
  (\bibinfo{year}{2021}).
\newblock
\newblock
\shownote{Electronic companion (full version) of the paper by the same name in
  Queueing Systems}.


\bibitem[Grosof et~al\mbox{.}(2022)]%
        {grosof_wcfs_2021}
\bibfield{author}{\bibinfo{person}{Isaac Grosof}, \bibinfo{person}{Mor
  Harchol-Balter}, {and} \bibinfo{person}{Alan Scheller-Wolf}.}
  \bibinfo{year}{2022}\natexlab{}.
\newblock \showarticletitle{WCFS: A new framework for analyzing multiserver
  systems}.
\newblock \bibinfo{journal}{\emph{Queueing Systems}} (\bibinfo{year}{2022}).
\newblock


\bibitem[Grosof et~al\mbox{.}(2018)]%
        {grosof_srpt_2018}
\bibfield{author}{\bibinfo{person}{Isaac Grosof}, \bibinfo{person}{Ziv Scully},
  {and} \bibinfo{person}{Mor Harchol-Balter}.} \bibinfo{year}{2018}\natexlab{}.
\newblock \showarticletitle{{SRPT} for multiserver systems}.
\newblock \bibinfo{journal}{\emph{Performance Evaluation}}
  \bibinfo{volume}{127-128} (\bibinfo{year}{2018}), \bibinfo{pages}{154--175}.
\newblock
\showISSN{0166-5316}


\bibitem[Grosof et~al\mbox{.}(2019)]%
        {grosof_guardrails_2019}
\bibfield{author}{\bibinfo{person}{Isaac Grosof}, \bibinfo{person}{Ziv Scully},
  {and} \bibinfo{person}{Mor Harchol-Balter}.} \bibinfo{year}{2019}\natexlab{}.
\newblock \showarticletitle{Load Balancing Guardrails: Keeping Your Heavy
  Traffic on the Road to Low Response Times}.
\newblock \bibinfo{journal}{\emph{Proc. ACM Meas. Anal. Comput. Syst.}}
  \bibinfo{volume}{3}, \bibinfo{number}{2}, Article \bibinfo{articleno}{42}
  (\bibinfo{date}{jun} \bibinfo{year}{2019}), \bibinfo{numpages}{31}~pages.
\newblock


\bibitem[{Guo} et~al\mbox{.}(2018)]%
        {guo_optimal_2018}
\bibfield{author}{\bibinfo{person}{Mian {Guo}}, \bibinfo{person}{Quansheng
  {Guan}}, {and} \bibinfo{person}{Wende {Ke}}.}
  \bibinfo{year}{2018}\natexlab{}.
\newblock \showarticletitle{Optimal Scheduling of VMs in Queueing Cloud
  Computing Systems With a Heterogeneous Workload}.
\newblock \bibinfo{journal}{\emph{IEEE Access}}  \bibinfo{volume}{6}
  (\bibinfo{year}{2018}), \bibinfo{pages}{15178--15191}.
\newblock


\bibitem[Harchol-Balter(2013)]%
        {harchol_performance_2013}
\bibfield{author}{\bibinfo{person}{Mor Harchol-Balter}.}
  \bibinfo{year}{2013}\natexlab{}.
\newblock \bibinfo{booktitle}{\emph{Performance modeling and design of computer
  systems: queueing theory in action}}.
\newblock \bibinfo{publisher}{Cambridge University Press}.
\newblock


\bibitem[Harchol-Balter(2022)]%
        {harchol_multiserver_2022}
\bibfield{author}{\bibinfo{person}{Mor Harchol-Balter}.}
  \bibinfo{year}{2022}\natexlab{}.
\newblock \showarticletitle{The multiserver job queueing model}.
\newblock \bibinfo{journal}{\emph{Queueing Systems}} \bibinfo{volume}{100},
  \bibinfo{number}{3} (\bibinfo{year}{2022}), \bibinfo{pages}{201--203}.
\newblock


\bibitem[Hong(2022)]%
        {hong_sharp_2022}
\bibfield{author}{\bibinfo{person}{Yige Hong}.}
  \bibinfo{year}{2022}\natexlab{}.
\newblock \showarticletitle{Sharp Zero-Queueing Bounds for Multi-Server Jobs}.
\newblock \bibinfo{journal}{\emph{SIGMETRICS Perform. Eval. Rev.}}
  \bibinfo{volume}{49}, \bibinfo{number}{2} (\bibinfo{date}{jan}
  \bibinfo{year}{2022}), \bibinfo{pages}{66–68}.
\newblock
\showISSN{0163-5999}


\bibitem[Institute(2020)]%
        {msi_queues}
\bibfield{author}{\bibinfo{person}{Minnesota~Supercomputing Institute}.}
  \bibinfo{year}{2020}\natexlab{}.
\newblock \bibinfo{title}{Queues}.
\newblock
\newblock
\urldef\tempurl%
\url{https://www.msi.umn.edu/queues}
\showURL{%
\tempurl}


\bibitem[Jones and Nitzberg(1999)]%
        {jones_scheduling}
\bibfield{author}{\bibinfo{person}{James~Patton Jones} {and}
  \bibinfo{person}{Bill Nitzberg}.} \bibinfo{year}{1999}\natexlab{}.
\newblock \showarticletitle{Scheduling for Parallel Supercomputing: A
  Historical Perspective of Achievable Utilization}. In
  \bibinfo{booktitle}{\emph{Job Scheduling Strategies for Parallel
  Processing}}, \bibfield{editor}{\bibinfo{person}{Dror~G. Feitelson} {and}
  \bibinfo{person}{Larry Rudolph}} (Eds.). \bibinfo{publisher}{Springer Berlin
  Heidelberg}, \bibinfo{address}{Berlin, Heidelberg}, \bibinfo{pages}{1--16}.
\newblock
\showISBNx{978-3-540-47954-3}


\bibitem[Kim(1979)]%
        {kim_mms_1979}
\bibfield{author}{\bibinfo{person}{Sung~Shick Kim}.}
  \bibinfo{year}{1979}\natexlab{}.
\newblock \emph{\bibinfo{title}{M/M/s queueing system where customers demand
  multiple server use}}.
\newblock \bibinfo{thesistype}{Ph.\,D. Dissertation}. \bibinfo{school}{Southern
  Methodist University}.
\newblock


\bibitem[Maguluri et~al\mbox{.}(2012)]%
        {maguluri_2012}
\bibfield{author}{\bibinfo{person}{Siva~Theja Maguluri},
  \bibinfo{person}{Rayadurgam Srikant}, {and} \bibinfo{person}{Lei Ying}.}
  \bibinfo{year}{2012}\natexlab{}.
\newblock \showarticletitle{Stochastic models of load balancing and scheduling
  in cloud computing clusters}. In \bibinfo{booktitle}{\emph{2012 Proceedings
  IEEE Infocom}}. IEEE, \bibinfo{pages}{702--710}.
\newblock


\bibitem[Psychas and Ghaderi(2018)]%
        {psychas_randomized_2018}
\bibfield{author}{\bibinfo{person}{Konstantinos Psychas} {and}
  \bibinfo{person}{Javad Ghaderi}.} \bibinfo{year}{2018}\natexlab{}.
\newblock \showarticletitle{Randomized Algorithms for Scheduling Multi-Resource
  Jobs in the Cloud}.
\newblock \bibinfo{journal}{\emph{IEEE/ACM Transactions on Networking}}
  \bibinfo{volume}{26}, \bibinfo{number}{5} (\bibinfo{year}{2018}),
  \bibinfo{pages}{2202--2215}.
\newblock


\bibitem[Rumyantsev et~al\mbox{.}(2022)]%
        {rumyantsevi_three_2022}
\bibfield{author}{\bibinfo{person}{Alexander Rumyantsev},
  \bibinfo{person}{Robert Basmadjian}, \bibinfo{person}{Sergey Astafiev}, {and}
  \bibinfo{person}{Alexander Golovin}.} \bibinfo{year}{2022}\natexlab{}.
\newblock \showarticletitle{Three-level modeling of a speed-scaling
  supercomputer}.
\newblock \bibinfo{journal}{\emph{Annals of Operations Research}}
  (\bibinfo{year}{2022}), \bibinfo{pages}{1--29}.
\newblock


\bibitem[Rumyantsev and Morozov(2017)]%
        {rumyantsev_2017}
\bibfield{author}{\bibinfo{person}{Alexander Rumyantsev} {and}
  \bibinfo{person}{Evsey Morozov}.} \bibinfo{year}{2017}\natexlab{}.
\newblock \showarticletitle{Stability criterion of a multiserver model with
  simultaneous service}.
\newblock \bibinfo{journal}{\emph{Annals of Operations Research}}
  \bibinfo{volume}{252}, \bibinfo{number}{1} (\bibinfo{year}{2017}),
  \bibinfo{pages}{29--39}.
\newblock


\bibitem[Schrage(1968)]%
        {schrage_proof_1968}
\bibfield{author}{\bibinfo{person}{Linus Schrage}.}
  \bibinfo{year}{1968}\natexlab{}.
\newblock \showarticletitle{A Proof of the Optimality of the Shortest Remaining
  Processing Time Discipline}.
\newblock \bibinfo{journal}{\emph{Operations Research}} \bibinfo{volume}{16},
  \bibinfo{number}{3} (\bibinfo{year}{1968}), \bibinfo{pages}{687--690}.
\newblock


\bibitem[Schrage and Miller(1966)]%
        {schrage_srpt_1966}
\bibfield{author}{\bibinfo{person}{Linus~E. Schrage} {and}
  \bibinfo{person}{Louis~W. Miller}.} \bibinfo{year}{1966}\natexlab{}.
\newblock \showarticletitle{The Queue {M/G/1} with the {Shortest Remaining
  Processing Time} Discipline}.
\newblock \bibinfo{journal}{\emph{Operations Research}} \bibinfo{volume}{14},
  \bibinfo{number}{4} (\bibinfo{year}{1966}), \bibinfo{pages}{670--684}.
\newblock


\bibitem[Scully(2021)]%
        {scully_wine_2021}
\bibfield{author}{\bibinfo{person}{Ziv Scully}.}
  \bibinfo{year}{2021}\natexlab{}.
\newblock \bibinfo{title}{{WINE}: A New Queueing Identity for Analyzing
  Scheduling Policies in Multiserver Systems}.
\newblock
\newblock
\urldef\tempurl%
\url{https://ziv.codes/pdf/wine-talk.pdf}
\showURL{%
\tempurl}
\newblock
\shownote{INFORMS Annual Meeting}.


\bibitem[Scully et~al\mbox{.}(2020)]%
        {scully_gittins_2020}
\bibfield{author}{\bibinfo{person}{Ziv Scully}, \bibinfo{person}{Isaac Grosof},
  {and} \bibinfo{person}{Mor Harchol-Balter}.} \bibinfo{year}{2020}\natexlab{}.
\newblock \showarticletitle{The {Gittins} Policy is Nearly Optimal in the
  {M/G/k} under Extremely General Conditions}.
\newblock \bibinfo{journal}{\emph{Proc. ACM Meas. Anal. Comput. Syst.}}
  \bibinfo{volume}{4}, \bibinfo{number}{3}, Article \bibinfo{articleno}{43}
  (\bibinfo{date}{Nov.} \bibinfo{year}{2020}), \bibinfo{numpages}{29}~pages.
\newblock


\bibitem[Scully et~al\mbox{.}(2021)]%
        {scully_optimal_2021}
\bibfield{author}{\bibinfo{person}{Ziv Scully}, \bibinfo{person}{Isaac Grosof},
  {and} \bibinfo{person}{Mor Harchol-Balter}.} \bibinfo{year}{2021}\natexlab{}.
\newblock \showarticletitle{Optimal multiserver scheduling with unknown job
  sizes in heavy traffic}.
\newblock \bibinfo{journal}{\emph{Performance Evaluation}}
  \bibinfo{volume}{145} (\bibinfo{year}{2021}), \bibinfo{pages}{102150}.
\newblock
\showISSN{0166-5316}


\bibitem[Scully and Harchol-Balter(2021)]%
        {scully_gittins_2021}
\bibfield{author}{\bibinfo{person}{Ziv Scully} {and} \bibinfo{person}{Mor
  Harchol-Balter}.} \bibinfo{year}{2021}\natexlab{}.
\newblock \showarticletitle{The Gittins Policy in the {M/G/1} Queue}. In
  \bibinfo{booktitle}{\emph{2021 19th International Symposium on Modeling and
  Optimization in Mobile, Ad hoc, and Wireless Networks (WiOpt)}}.
  \bibinfo{pages}{1--8}.
\newblock


\bibitem[Srinivasan et~al\mbox{.}(2002)]%
        {srinivasan_characterization_2002}
\bibfield{author}{\bibinfo{person}{Srividya Srinivasan},
  \bibinfo{person}{Rajkumar Kettimuthu}, \bibinfo{person}{Vijay Subramani},
  {and} \bibinfo{person}{Ponnuswamy Sadayappan}.}
  \bibinfo{year}{2002}\natexlab{}.
\newblock \showarticletitle{Characterization of backfilling strategies for
  parallel job scheduling}. In \bibinfo{booktitle}{\emph{Proceedings.
  International Conference on Parallel Processing Workshop}}.
  \bibinfo{pages}{514--519}.
\newblock


\bibitem[{Tang} et~al\mbox{.}(2011)]%
        {tang_reducing}
\bibfield{author}{\bibinfo{person}{Wei {Tang}}, \bibinfo{person}{Zhiling
  {Lan}}, \bibinfo{person}{Narayan {Desai}}, \bibinfo{person}{Daniel
  {Buettner}}, {and} \bibinfo{person}{Yongen {Yu}}.}
  \bibinfo{year}{2011}\natexlab{}.
\newblock \showarticletitle{Reducing Fragmentation on Torus-Connected
  Supercomputers}. In \bibinfo{booktitle}{\emph{2011 IEEE International
  Parallel Distributed Processing Symposium}}. \bibinfo{pages}{828--839}.
\newblock


\bibitem[{Tang} et~al\mbox{.}(2012)]%
        {tang_adaptive}
\bibfield{author}{\bibinfo{person}{Wei {Tang}}, \bibinfo{person}{Dongxu {Ren}},
  \bibinfo{person}{Zhiling {Lan}}, {and} \bibinfo{person}{Narayan {Desai}}.}
  \bibinfo{year}{2012}\natexlab{}.
\newblock \showarticletitle{Adaptive Metric-Aware Job Scheduling for Production
  Supercomputers}. In \bibinfo{booktitle}{\emph{2012 41st International
  Conference on Parallel Processing Workshops}}. \bibinfo{pages}{107--115}.
\newblock


\bibitem[Tikhonenko(2005)]%
        {tikhonenko_generalized_2005}
\bibfield{author}{\bibinfo{person}{Oleg~M. Tikhonenko}.}
  \bibinfo{year}{2005}\natexlab{}.
\newblock \showarticletitle{Generalized Erlang problem for service systems with
  finite total capacity}.
\newblock \bibinfo{journal}{\emph{Problems of Information Transmission}}
  \bibinfo{volume}{41}, \bibinfo{number}{3} (\bibinfo{year}{2005}),
  \bibinfo{pages}{243--253}.
\newblock


\bibitem[Tirmazi et~al\mbox{.}(2020)]%
        {tirmazi_2020}
\bibfield{author}{\bibinfo{person}{Muhammad Tirmazi}, \bibinfo{person}{Adam
  Barker}, \bibinfo{person}{Nan Deng}, \bibinfo{person}{Md~E. Haque},
  \bibinfo{person}{Zhijing~Gene Qin}, \bibinfo{person}{Steven Hand},
  \bibinfo{person}{Mor Harchol-Balter}, {and} \bibinfo{person}{John Wilkes}.}
  \bibinfo{year}{2020}\natexlab{}.
\newblock \showarticletitle{Borg: The next Generation}. In
  \bibinfo{booktitle}{\emph{Proceedings of the Fifteenth European Conference on
  Computer Systems}} (Heraklion, Greece) \emph{(\bibinfo{series}{EuroSys
  '20})}. \bibinfo{publisher}{Association for Computing Machinery},
  \bibinfo{address}{New York, NY, USA}, Article \bibinfo{articleno}{30},
  \bibinfo{numpages}{14}~pages.
\newblock
\showISBNx{9781450368827}


\bibitem[van Dijk(1989)]%
        {vandijk_blocking_1989}
\bibfield{author}{\bibinfo{person}{Nico~M. van Dijk}.}
  \bibinfo{year}{1989}\natexlab{}.
\newblock \showarticletitle{Blocking of finite source inputs which require
  simultaneous servers with general think and holding times}.
\newblock \bibinfo{journal}{\emph{Operations Research Letters}}
  \bibinfo{volume}{8}, \bibinfo{number}{1} (\bibinfo{year}{1989}),
  \bibinfo{pages}{45 -- 52}.
\newblock
\showISSN{0167-6377}


\bibitem[Vizino et~al\mbox{.}(2005)]%
        {vizino_batch}
\bibfield{author}{\bibinfo{person}{Chad Vizino}, \bibinfo{person}{Nathan
  Stone}, \bibinfo{person}{John Kochmar}, {and} \bibinfo{person}{J.~Ray
  Scott}.} \bibinfo{year}{2005}\natexlab{}.
\newblock \showarticletitle{Batch Scheduling on the Cray XT3}.
\newblock \bibinfo{journal}{\emph{CUG 2005}} (\bibinfo{year}{2005}).
\newblock


\bibitem[Wang and Guo(2009)]%
        {wang_application_2009}
\bibfield{author}{\bibinfo{person}{Juan Wang} {and} \bibinfo{person}{Wenming
  Guo}.} \bibinfo{year}{2009}\natexlab{}.
\newblock \showarticletitle{The Application of Backfilling in Cluster Systems}.
  In \bibinfo{booktitle}{\emph{2009 WRI International Conference on
  Communications and Mobile Computing}}, Vol.~\bibinfo{volume}{3}.
  \bibinfo{pages}{55--59}.
\newblock


\bibitem[Wang et~al\mbox{.}(2021)]%
        {wang_zero_2021}
\bibfield{author}{\bibinfo{person}{Weina Wang}, \bibinfo{person}{Qiaomin Xie},
  {and} \bibinfo{person}{Mor Harchol-Balter}.} \bibinfo{year}{2021}\natexlab{}.
\newblock \showarticletitle{Zero Queueing for Multi-Server Jobs}. In
  \bibinfo{booktitle}{\emph{Abstract Proceedings of the 2021 ACM SIGMETRICS /
  International Conference on Measurement and Modeling of Computer Systems}}
  (Virtual Event, China) \emph{(\bibinfo{series}{SIGMETRICS '21})}.
  \bibinfo{publisher}{Association for Computing Machinery},
  \bibinfo{address}{New York, NY, USA}, \bibinfo{pages}{13–14}.
\newblock
\showISBNx{9781450380720}


\bibitem[Whitt(1985)]%
        {whitt_blocking_1985}
\bibfield{author}{\bibinfo{person}{Ward Whitt}.}
  \bibinfo{year}{1985}\natexlab{}.
\newblock \showarticletitle{Blocking when service is required from several
  facilities simultaneously}.
\newblock \bibinfo{journal}{\emph{AT\&T Technical Journal}}
  \bibinfo{volume}{64}, \bibinfo{number}{8} (\bibinfo{year}{1985}),
  \bibinfo{pages}{1807--1856}.
\newblock


\end{thebibliography}

\appendix
\section{Proof of Lemma~\ref{lem:decomp} (Work decomposition)}
\label{app:work-decomp}
\begin{replemma}{lem:decomp}{\cite[Theorem~7.2]{scully_gittins_2020}}
    For an arbitrary scheduling policy $\pi$,
    in an arbitrary system,
    \begin{align*}
        E[W^\pi_r] - E[W^{SRPT \hy 1}_r]
        = \frac{E[(1-B^\pi_r) W^\pi_r] + \rho_r^R E_r[W_r^\pi]}{1-\rho_r^A}.
    \end{align*}
\end{replemma}
\begin{proof}
    We will employ the rate conservation law, applied to the random variable
    $(W_r^\pi)^2$,
    the square of the stationary distribution of $r$-relevant work in the system.
    The rate conservation law
    states that, because $(W_r^\pi)^2$ is a stationary random variable,
    its expected rate of increase and decrease must be equal.
    This argument can be formalized further using Palm Calculus.

    To find these rates of increase and decrease, let us first examine $W_r^\pi$.
    $W_r^\pi$ decreases continuously as work completes,
    and increases by jumps whenever jobs arrive.
    $W_r^\pi$ decreases at rate $B_r^\pi$,
    the fraction of servers that are occupied by $r$-relevant jobs.
    When a job arrives with size $S$, it contributes $[S \mathds{1}\{S \le r\}]$
    relevant work, increasing $W_r^\pi$ by that amount.
    Such arrivals occur at rate $\lambda$.
    Finally, whenever a job recycles, by being served until its remaining size
    falls to $r$, it adds $r$ relevant work to $W_r^\pi$.

    Using these rates, we can calculate the expected rates of increase and decrease of $(W_r^\pi)^2$.
    \begin{align*}
        \text{Increase due to arrivals:} \quad &\lambda E[(S \mathds{1}\{S \le r\})^2] + 2\rho_r^A E[W_r^\pi] \\
        \text{Increase due to recycling:} \quad &\lambda^R_r r^2 + 2 \rho_r^R E_r[W_r^\pi] \\
        \text{Decrease due to service:} \quad &2 E[B_r^\pi W_r^\pi]
    \end{align*}
    Equating these rates, we find that
    \begin{align}
    \nonumber
        2 E[B_r^\pi W_r^\pi]
        &= \lambda E[(S \mathds{1}\{S \le r\})^2] + 2\rho_r^A E[W_r^\pi]
        + \lambda^R_r r^2 + 2 \rho_r^R E_r[W_r^\pi]. \\
    \nonumber
        E[B_r^\pi W_r^\pi] 
        &= \frac{\lambda}{2} E[(S \mathds{1}\{S \le r\})^2]
        + \rho_r^A E[W_r^\pi]
        + \frac{\lambda^R_r}{2} r^2
        + \rho_r^R E_r[W_r^\pi]. \\
    \nonumber
        E[W_r^\pi] - E[(1-B_r^\pi) W_r^\pi]
                &= \frac{\lambda}{2} E[(S \mathds{1}\{S \le r\})^2]
        + \rho_r^A E[W_r^\pi]
        + \frac{\lambda^R_r}{2} r^2
        + \rho_r^R E_r[W_r^\pi]. \\
    \nonumber
        E[W_r^\pi]
            &=
          E[(1-B_r^\pi) W_r^\pi]
        + \frac{\lambda}{2} E[(S \mathds{1}\{S \le r\})^2]
        + \rho_r^A E[W_r^\pi]
        + \frac{\lambda^R_r}{2} r^2
        + \rho_r^R E_r[W_r^\pi]. \\
    \nonumber
        E[W_r^\pi] (1-\rho_r^A)
            &=
          E[(1-B_r^\pi) W_r^\pi]
        + \frac{\lambda}{2} E[(S \mathds{1}\{S \le r\})^2]
        + \frac{\lambda^R_r}{2} r^2
        + \rho_r^R E_r[W_r^\pi]. \\
    \label{eq:main-decomp}
        E[W_r^\pi] (1-\rho_r^A)
            &=
          E[(1-B_r^\pi) W_r^\pi]
        + \rho_r^R E_r[W_r^\pi]
        + \frac{\lambda}{2} E[(S \mathds{1}\{S \le r\})^2]
        + \frac{\lambda^R_r}{2} r^2.
    \end{align}
    Let us evaluate \eqref{eq:main-decomp} in the case where the policy $\pi$ is SRPT-1.
    The first two terms of the right-hand side are nonnegative terms depending on the policy $\pi$,
    while the second two terms are the same for all policies.

    Let us start with the first term on the right-hand side,
    $E[(1-B_r^\pi) W_r^\pi]$.
    Note that under SRPT-1, if $W_r^\pi$ is nonzero,
    i.e. if a $r$-relevant job is present,
    then SRPT-1 will serve a $r$-relevant job on its single server,
    and so $B_r^{SRPT \hy 1} = 1$.
    As a result, either $W_r^{SRPT \hy 1}$ or $1-B_r^{SRPT \hy 1}$ must always be zero,
    so this term is equal to 0.

    Next, consider the second term, $\rho_r^R E_r[W_r^\pi]$.
    Recall that $E_r[\cdot]$ is an expectation over system states at times when
    $r$-relevant jobs recycle.
    In the SRPT-1 system, if a job is recycling by falling down to remaining size $r$,
    there must be no jobs in the system with remaining size less than $r$.
    As a result, $E_r[W_r^\pi] = 0$.

    We therefore conclude that
    \begin{align}
        \label{eq:srpt-decomp}
        E[W_r^{SRPT \hy 1}] (1-\rho_r^A)
        = \frac{\lambda}{2} E[(S \mathds{1}\{S \le r\})^2] + \frac{\lambda^R_r}{2} r^2.
    \end{align}
    As an aside, note that this argument shows that SRPT-1 has the least value of $E[W_r^\pi]$
    for any policy $\pi$. This fact, combined with \cref{lem:wine},
    provides an alternative proof that SRPT-1 is the optimal scheduling policy in the M/G/1.

    Subtracting \eqref{eq:srpt-decomp} from \eqref{eq:main-decomp},
    we find that
    \begin{align*}
        E[W_r^\pi] (1-\rho_r^A) - E[W_r^{SRPT \hy 1}] (1-\rho_r^A)
        &= E[(1-B_r^\pi) W_r^\pi] + \rho_r^R E_r[W_r^\pi] \\
        E[W^\pi_r] - E[W^{SRPT \hy 1}_r]
        &= \frac{E[(1-B^\pi_r) W^\pi_r] + \rho_r^R E_r[W_r^\pi]}{1-\rho_r^A}.
    \end{align*}
\end{proof}

\section{ServerFilling-Gittins proofs}
\label{app:gittins}

Our results for ServerFilling-Gittins follow near-identical proofs as given in \cref{sec:proof}
for ServerFilling-SRPT. We give the proofs here for completeness.

Our starting point is the ``work integral number equality'' (WINE) identity \cite{scully_gittins_2020,scully_wine_2021}.
\begin{theorem}[Theorem 6.3, \cite{scully_gittins_2020}]
    \label{thm:wine-gittins}
    The mean number of jobs and mean response time in an arbitrary system,
    under an arbitrary scheduling policy,
    is
    \begin{align*}
        E[N] = \lambda E[T] = \int_0^\infty \frac{E[W_r]}{r^2} dr.
    \end{align*}
\end{theorem}

Now, we can state the work-decomposition law in a Gittins system.
\begin{theorem}[Theorem 7.2, \cite{scully_gittins_2020}]
    \label{thm:decomp-gittins}
    For all $r \ge 0$,
    the mean $r$-relevant work gap between an arbitrary policy $\pi$ and M/G/1/Gittins is
    \begin{align}
        \label{eq:decomp-gittins}
        E[W^\pi_r] - E[W^{Gittins \hy 1}_r] = \frac{E[(1-B^\pi_r)W^\pi_r]
        + \lambda^R_r E_r[S_r(X^R_r)W^\pi_r]}{1-\rho^A_{r}}.
    \end{align}
\end{theorem}

We will handle the two numerator terms of \eqref{eq:decomp-gittins} separately.
Let us start by combining \cref{thm:wine-gittins} with \cref{thm:decomp-gittins},
and try to bound the resulting integral.

We must bound
\begin{align*}
    E[T^\pi] - E[T^{Gittins \hy 1}] = \frac{1}{\lambda} \int_0^\infty \frac{E[(1-B^\pi_r)W^\pi_r]}{r^2(1-\rho^A_{r})}
    + \frac{1}{\lambda} \int_0^\infty \frac{\lambda^R_r E_r[S_r(X^R_r)W^\pi_r]}{r^2(1-\rho^A_{r})}.
\end{align*}

We bound the first term in \cref{lem:bound-idle-gittins} and the second term in \cref{lem:bound-recycle-gittins}.

\begin{lemma}
    \label{lem:bound-idle-gittins}
    \begin{align*}
        \int_{r=0}^\infty \frac{E[(1-B_r) W_r]}{r^2(1-\rho^A_{r})} dr
        \le e(k-1)\lceil \ln \frac{1}{1-\rho} \rceil.
    \end{align*}
\end{lemma}
\begin{proof}
    First, we make use of the key fact about ServerFilling-Gittins (and DivisorFilling-Gittins):
    If there are at least $k$ jobs with rank $\le r$ in the system,
    then $B_r = 1$.
    Thus, we can replace $W_r$ by $W'_r$,
    the work of the $k-1$ jobs of least rank in the system:
        \begin{align*}
        \int_{r=0}^\infty \frac{E[(1-B_r) W_r]}{r^2(1-\rho^A_{r})} dr =
        \int_{r=0}^\infty \frac{E[(1-B_r) W'_r]}{r^2(1-\rho^A_{r})} dr.
    \end{align*}

    Next, we will break up the ranks $r \in [0, \infty)$ into a finite set of buckets.
    Let $R = [r_1, r_2, \ldots]$ be a list of ranks, where $r_1 = 0$.
    We will specify the list $R$ later.
    Implicitly, we will say that $r_{|R|+1} = \infty$.
    We can rewrite the above integral as:
    \begin{align}
        \label{eq:bucketed-gittins}
        \int_{r=0}^\infty \frac{E[(1-B_r) W'_r]}{r^2(1-\rho^A_{r})} dr =
        \sum_{i=1}^{|R|} \int_{r=r_i}^{r_{i+1}} \frac{E[(1-B_r) W'_r]}{r^2(1-\rho^A_{r})} dr.
    \end{align}
    Next, we replace $r$ with either $r_i$ or $r_{i+1}$, selectively, to simplify things.
    Note that $B_r$ is increasing as a function of $r$ - as we increase the rank $r$,
    more servers are busy with $r$-relevant jobs.
    Likewise, $\rho^A_{r}$ is increasing as a function of $r$.
    Thus,
    \begin{align*}
        B_{r_i} &\le B_r \\
        \rho^A_{r} &\le \rho^A_{r_{i+1}}.
    \end{align*}
    Substituting into the integral from \eqref{eq:bucketed},
    we find that
    \begin{align*}
        \int_{r=r_i}^{r_{i+1}} \frac{E[(1-B_r) W'_r]}{r^2(1-\rho^A_{r})} dr
        \le \int_{r=r_i}^{r_{i+1}} \frac{E[(1-B_{r_i}) W'_r]}{r^2(1-\rho^A_{r_{i+1}})} dr.
    \end{align*}
    Next, let us perform some algebraic manipulation:
    \begin{align*}
        \int_{r=r_i}^{r_{i+1}} \frac{E[(1-B_{r_i}) W'_r]}{r^2(1-\rho^A_{r_{i+1}})} dr
        =E \left[ \int_{r=r_i}^{r_{i+1}}
            \frac{(1-B_{r_i}) W'_r}{r^2(1-\rho^A_{r_{i+1}})} dr \right]
        =E \left[ \frac{1-B_{r_i}}{1-\rho^A_{r_{i+1}}} 
        \int_{r=r_i}^{r_{i+1}} \frac{W'_r}{r^2} dr \right].
    \end{align*}
    Note that $B_{r_i}$ and $W'_r$ are conditionally independent
    because given $\vec{X}$, the current states of the jobs in the system,
    the busyness $B_{r_i}$ is deterministic.
    We can make this explicit:
    \begin{align}
        \label{eq:cond-ind-gittins}
        E \left[ \frac{1-B_{r_i}}{1-\rho^A_{r_{i+1}}} \int_{r=r_i}^{r_{i+1}} \frac{W'_r}{r^2} \right] dr
        &= E \left[ \frac{1-B_{r_i}}{1-\rho^A_{r_{i+1}}}
    \int_{r=r_i}^{r_{i+1}} \frac{E[W'_r \mid \vec{X}]}{r^2} dr \right].
    \end{align}
    Next, let us recall the definition of $W'_r$:
    \begin{align*}
        W'_r &= \sum_{j=1}^{k-1} S_r(X_j), \\
        E[W'_r \mid \vec{X}] &= \sum_{j=1}^{k-1} E[S_r(X_j) | X_j].
    \end{align*}
    Following \cite{scully_gittins_2020}, let us define
    $\service(X_j, r)$ to be $E[S_r(X_j) | X_j]$,
    the expected $r$-relevant work of a job $X_j$.

    Substituting this into \eqref{eq:cond-ind-gittins},
    we find that
    \begin{align}
        \nonumber
        &E \left[ \frac{1-B_{r_i}}{1-\rho^A_{r_{i+1}}}
        \int_{r=r_i}^{r_{i+1}} \frac{E[W'_r \mid \vec{X}]}{r^2} dr \right] \\
        \label{eq:service-gittins}
        = &E \left[ \frac{1-B_{r_i}}{1-\rho^A_{r_{i+1}}}
            \sum_{j=1}^{k-1} \int_{r=r_i}^{r_{i+1}} \frac{\service(X_j, r)}{r^2} dr \right].
    \end{align}
    Now, let us make use of the basic fact about $\service(X_j, r)$ from \cite{scully_gittins_2020}
    which underlies \cref{thm:wine-gittins}:

    For any job state $X_j$ which is not the empty job,
    \begin{align*}
        \int_{r=0}^\infty \frac{\service(X_j, r)}{r^2} dr = 1.
    \end{align*}
    For the empty job, service is 0.

    This provides a loose bound on the integral in \eqref{eq:service},
    which integrates over a smaller interval of ranks. Substituting in this bound,
    we find that
    \begin{align*}
        &E \left[ \frac{1-B_{r_i}}{1-\rho^A_{r_{i+1}}}
        \sum_{j=1}^{k-1} \int_{r=r_i}^{r_{i+1}} \frac{\service(X_j, r)}{r^2} dr \right]
        \le E \left[ \frac{1-B_{r_i}}{1-\rho^A_{r_{i+1}}} \min \{N, k-1 \} \right] \\
        &\le (k-1) E \left[ \frac{1-B_{r_i}}{1-\rho^A_{r_{i+1}}} \right]
        = (k-1) \frac{1-\rho^A_{r_i} - \rho^R_r}{1-\rho^A_{r_{i+1}}}
        \le (k-1) \frac{1-\rho^A_{r_i}}{1-\rho^A_{r_{i+1}}}.
    \end{align*}
    Returning all the way back to the beginning, we find that
    \begin{align*}
        \int_{r=0}^\infty \frac{E[(1-B_r) W_r]}{r^2(1-\rho^A_{r})} dr
        \le (k-1)\sum_{r=0}^{|R|} \frac{1-\rho^A_{r_i}}{1-\rho^A_{r_{i+1}}}.
    \end{align*}
    To optimize this bound, we need to choose $R$ to minimize this sum.
    To do so, we set $|R| = \lceil \ln \frac{1}{1-\rho} \rceil$,
    and choose $r_i$ such that
    \begin{align*}
        \frac{1-\rho^A_{r_i^+}}{1-\rho^A_{r_{i+1}^-}} \le e.
    \end{align*}
    for all $i < |R|$.
    By $^+$ and $^-$, we refer to the left and right limits,
    thereby handling the possibility that $\rho^A_r$ is discontinuous as a function of $r$.
    We therefore find that
    \begin{align*}
        \int_{r=0}^\infty \frac{E[(1-B_r) W_r]}{r^2(1-\rho^A_{r})} dr
        \le e(k-1) \biggl \lceil \ln \frac{1}{1-\rho} \biggr \rceil.
    \end{align*}
\end{proof}
Now, it remains to bound the recyclings term in \eqref{eq:decomp-gittins}.
Note that this term is identical to the one in \cite{scully_gittins_2020},
so we can use essentially the same approach - we just disentangle it from the other term.
First, we use a basic theorem from \cite{scully_gittins_2020}:
\begin{lemma}[Lemma 8.2, \cite{scully_gittins_2020}]
    \begin{align*}
        \lambda^R_r E_r[S_r(X^R_r) W_r] \le (k-1)r \rho^R_r.
    \end{align*}
\end{lemma}

Now, it remains to bound the recyclings-dependent term, plugged into \cref{thm:wine-gittins}.
\begin{lemma}
    \label{lem:bound-recycle-gittins}
    \begin{align*}
        \int_{r=0}^\infty \frac{(k-1)r \rho^R_r}{r^2(1-\rho^A_{r})} dr
        \le (k-1) \ln \frac{1}{1-\rho}
    \end{align*}
\end{lemma}
\begin{proof}
    First, let us simplify:
    \begin{align*}
        \int_{r=0}^\infty \frac{(k-1)r \rho^R_r}{r^2(1-\rho^A_{r})} dr
        = (k-1) \int_{r=0}^\infty \frac{\rho^R_r}{1-\rho^A_{r}} \frac{1}{r} dr.
    \end{align*}
    To bound the integrand, we will explicitly consider the Gittins game.
    Using the definitions of $\undone_A(r)$, and $\game_A(r)$
    given in Appendix B.2 of \cite{scully_gittins_2020},
    we bound as follows:
    \begin{align*}
        \frac{\rho^R_r}{1-\rho^A_{r}}
        \frac{1}{r}
        &\le
        \frac{\lambda r \undone_A(r)}{1 - \lambda (\game_A(r) - r \undone_A(r))}
        \frac{1}{r} \\
        &\le \frac{\lambda \undone_A(r)}{1 - \lambda \game_A(r)}
        = \frac{\lambda \frac{d}{dr} \game_A(r)}{1 - \lambda \game_A(r)}
        = \frac{d}{dr} \ln \frac{1}{1-\lambda \game_A(r)}.
    \end{align*}
    Above, we make use of \cite[Lemma 5.3]{scully_gittins_2020}.

    Integrating over all $r \in [0, \infty)$, we find that
    \begin{align*}
        \int_{r=0}^\infty \frac{\rho^R_r}{1-\rho^A_{r}} \frac{1}{r} dr
        &\le \left[ \ln \frac{1}{1 - \lambda \game_A(r)} \right]_{r=0}^\infty \\
        &= \ln \frac{1}{1-\lambda \game_A(\infty)}
        - \ln \frac{1}{1 - \lambda \game_A(0)}.
    \end{align*}
    From the definition of the Gittins game,
    it is straightforward to prove that
    $\game_A(0) = 0$, and that $\game_A(\infty) = E[S]$.

    As a result,
    \begin{align*}
        \int_{r=0}^\infty \frac{\rho^R_r}{1-\rho^A_{r}} \frac{1}{r} dr
        \le \ln \frac{1}{1-\rho}.
    \end{align*}
\end{proof}

Now, we're ready to put it all together.
We derive a bound on mean response time:
\begin{reptheorem}{thm:gittins-bound}
    In any multiserver-job system
    in the power-of-two setting
    the difference in mean response time between ServerFilling-Gittins
    and Gittins-1 (resource pooled) is at most
    \begin{align*}
        E[T^{SFG \hy k}] - E[T^{Gittins \hy 1}]
            \le  \frac{(e+1)(k-1)}{\lambda} \ln \left( \frac{1}{1-\rho} \right) + \frac{e}{\lambda}.
    \end{align*}
    The same is true of DivisorFilling-Gittins in the divisible setting.
\end{reptheorem}
\begin{proof}
    Combine \cref{thm:wine-gittins} with \cref{thm:decomp-gittins}, using \cref{lem:bound-idle-gittins} and \cref{lem:bound-recycle-gittins} to bound the two terms.
\end{proof}
Note that this bound is in some ways stronger than the bound on Gittins-$k$
given in \cite{scully_gittins_2020}.
Our bound is the first uniform bound on multiserver Gittins,
meaning that our bound doesn't depend on $S$ except via $E[S]$,
unlike the bound on Gittins-$k$ in \cite{scully_gittins_2020}.
Note also that the M/G/$k$ is a special case of the multiserver-job system when server needs are all 1,
and that in this special case, ServerFilling-Gittins specializes to Gittins-$k$.
As a result, \cref{thm:gittins-bound} is a strict improvement
upon the bound given in \cite{scully_gittins_2020}.

Analogous to \cref{thm:sfs-opt},
we use our bound to prove that ServerFilling-Gittins (and DivisorFilling-Gittins)
achieve asymptotically optimal mean response time.

\begin{reptheorem}{thm:gittins-opt}
    If $E[S^2(\log S)^+] < \infty$,
    \begin{align*}
        \lim_{\rho \to 1} \frac{E[T^{SFG \hy k}]}{E[T^{Gittins \hy 1}]} = 1.
    \end{align*}
\end{reptheorem}
Note that $E[T^{SRPT \hy 1}] \le E[T^{Gittins \hy 1}]$ by the optimality of SRPT,
so $E[T^{Gittins \hy 1}] = \omega(\frac{1}{1-\rho})$ whenever $E[S^2(\log S)^+] < \infty$,
just as $E[T^{SRPT \hy 1}] = \omega(\frac{1}{1-\rho})$ in this case.

\section{DivisorFilling-SRPT}
\label{app:divisor-filling}
The DivisorFilling-SRPT policy is a scheduling policy for the divisible server needs setting
of the multiserver-job system,
where all server needs $k_j$ perfectly divide the total number of servers $k$.

To implement DivisorFilling-SRPT, we order jobs in increasing order of remaining size $r_j$,
and then apply a recursive procedure to select the jobs to serve,
which we will specify in \cref{sec:dfs-definition}.
DivisorFilling-Gittins is defined identically,
replacing increasing remaining size order with increasing rank order.

DivisorFilling-SRPT achieves two key guarantees:
\begin{enumerate}
\item \label{item:trivial} DivisorFilling-SRPT always serves a subset of the $k$ jobs of least remaining size in the system.
\item If at least $k$ jobs are present, DivisorFilling-SRPT serves jobs with total server need exactly $k$.
\label{item:hard}
\end{enumerate}
\cref{item:trivial} is part of the definition of DivisorFilling-SRPT
in \cref{sec:dfs-definition}.
We prove \cref{item:hard} as \cref{thm:dfs-proof} in \cref{sec:dfs-proof}.

DivisorFilling-SRPT is identical to the DivisorFilling policy
defined in \cite[Appendix A]{grosof_wcfs_2021_full},
except that DivisorFilling orders jobs in the arrival ordering,
while DivisorFilling-SRPT orders jobs in SRPT order.
Note that \cite{grosof_wcfs_2021_full} is the electronic companion to
\cite{grosof_wcfs_2021}, and that Appendix A only appears in the electronic companion.

As a corollary of \cref{item:trivial,item:hard},
we can prove the ``relevant work efficiency'' property for DivisorFilling-SRPT:
\begin{corollary}[Relevant work efficiency]
    \label{cor:dfs-rwe}
    Under the DivisorFilling-SRPT policy,
    in the divisible setting,
    if there are $k$ or more $r$-relevant jobs in the system,
    all servers are occupied by $r$-relevant jobs.

    The same is true for DivisorFilling-Gittins.
\end{corollary}
From \cref{cor:dfs-rwe},
we can use the same techniques as were used for ServerFilling-SRPT
to prove \cref{thm:sfs-bound,thm:sfs-opt}.

\subsection{DivisorFilling-SRPT Definition}
\label{sec:dfs-definition}

Order all jobs in the system in order of least remaining size.
Let $M$ be the set of $k$ jobs with least remaining size,
or all jobs if less than $k$ are present.

We now split into three cases:
\begin{enumerate}
    \item $M$ contains at least $k/6$ jobs with server need $k_j = 1$.
    \item $k = 2^a3^b$ for some integers $a, b$, and $M$ contains $< k/6$ jobs with $k_j = 1$.
    \item $k$ has largest prime factor $p \ge 5$, and $M$ contains $< k/6$ jobs with $k_j = 1$.
\end{enumerate}

\subsubsection{Case 1}
If $M$ contains at least $k/6$ jobs with server need 1,
we initially parallel the ServerFilling-SRPT policy:
we order jobs in $M$ by server need (tiebroken by least remaining size),
and place jobs into service in that order.
However, because server needs are not powers of two,
we may reach a point where no more jobs fit into service, but servers are still unoccupied.
In this case, we place jobs from $M$ with server need 1 into service,
again tiebroken by least remaining size.
We continue doing so until all $k$ servers are full or no more server need 1 jobs remain.

\subsubsection{Case 2}
Suppose that $k$ is of the form $2^a3^b$,
and that Case 1 does not apply.

We will recurse on one of two subsets of $M$: the set of jobs with even server need,
or the set of jobs of odd server need greater than 1.
Note that all jobs in the latter subset have server needs divisible by 3.
We call the former subset $M_2$ and the latter subset $M_3$.
To decide which subset to recurse on,
we compare the values $2|M_2|$ and $3|M_3|$,
and recurse on the subset whose value is larger.
In the case of a tie, we arbitrarily select $M_2$.

If $2|M_2|$ is larger, we will only serve jobs from among $M_2$.
To decide which jobs to serve,
imagine that we combine pairs of servers.
Doing so reduces $k$ by a factor of 2,
and reduces the server need of each job in $M_2$ by a factor of 2.
We now recursively compute which jobs from $M_2$
the DivisorFilling-SRPT policy would serve in this subproblem,
and serve those same jobs.
If $3|M_3|$ is larger, we combine triples of servers, and then perform the same recursion.

\subsubsection{Case 3}
Suppose that $k$ has largest prime factor $p \ge 5$,
and that Case 1 does not apply.

Let $M_p$ be the set of jobs in $M$ with server need divisible by $p$.
If $p|M_p| \ge k$, we recurse as in Case~2 by combining groups of $p$ servers.

Otherwise, we will only serve jobs from $M$ whose server need is \emph{not} divisible by $p$,
and also greater than 1.
Let $M_r$ be this subset of $M$.
Note that all jobs in $M_r$ have server needs which are divisors of $k/p$.
We therefore construct a set $M'$ consisting of the $k/p$ jobs of $M_r$
with least remaining size.
If less than $k/p$ jobs are in $M_r$, $M'$ is all of $M_r$.
We then apply the DivisorFilling-SRPT procedure to $M'$,
setting the total number of servers $k' = k/p$ in the subproblem.
We extract the subset of jobs that DivisorFilling-SRPT serves in the subproblem from
$M_r$.
We repeat this process by extracting subsets from the remaining jobs in $M_r$,
repeating until we have extracted $p$ subsets from $M_r$, or $M_r$ contains no jobs.
DivisorFilling-SRPT serves all jobs that were served in any of the $p$ subproblems.

Note that this set of jobs served is valid to serve,
with total server need at most $k$,
because each of the $p$ subproblems have total server need at most $k/p$.

\subsection{DivisorFilling-SRPT Fills All Servers}
\label{sec:dfs-proof}

Our proof mirrors the proof in \cite[Appendix~A]{grosof_wcfs_2021_full},
which we reprove to make this paper self-contained.

\begin{theorem}
    \label{thm:dfs-proof}
    If at least $k$ jobs are present, DivisorFilling-SRPT
    serves a set of jobs with total server need exactly $k$.

    The same is true for DivisorFilling-Gittins.
\end{theorem}
\begin{proof}
    We will prove that if $M$ contains $k$ jobs,
    DivisorFilling-SRPT serves all $k$ jobs.
    Our proof proceeds by strong induction on $k$.
    Specifically, assume that for all $k' < k$,
    if $M'$ consists of at least $k'$ jobs whose server needs divide $k'$,
    then DivisorFilling-SRPT run on $M'$ serves a set of jobs with total server need $k'$.
    We will show that this assumption implies the desired result for $k$ servers.

    Again, we split into three cases:
    \begin{enumerate}
        \item $M$ contains at least $k/6$ jobs with server need $k_j$=1.
        \item $k=2^a3^b$ for some integers $a, b,$ and $M$ contains $<k/6$ jobs with $k_j=1$.
        \item $k$ has largest prime factor $p \ge 5$, and $M$ contains $<k/6$ jobs with $k_j=1$.
    \end{enumerate}
\subsubsection{Case 1} Suppose $M$ contains at least $k/6$ jobs with server need 1.

Let us label the jobs in $M$ as $m_1, m_2, \ldots$
in decreasing order of server need:
\begin{align*}
    k_{m_1} \ge k_{m_2} \ge \ldots.
\end{align*}
Let $i^*$ be defined as
\begin{align*}
    i^* = \arg \max_i \sum_{\ell=1}^i k_{m_\ell} \le k.
\end{align*}
In Case 1, DivisorFilling-SRPT serves jobs $m_1, \ldots m_{i^*}$,
as well as any jobs with $k_j = 1$ that fit in the remaining servers.
Let us write $\sumsc_i := \sum_{\ell=1}^{i} k_{m_\ell}$.
Because $M$ contains at least $k/6$ jobs with server need 1,
to prove \cref{thm:dfs-proof} in this case,
it suffices to show that $\sumsc_{i^*} \ge 5k/6$.
The remaining servers are filled by the jobs with server need 1.

First, note that $\sumsc_k \ge k$, because $M$ contains $k$ jobs,
each with server need at least 1.
Next, note that $k - \sumsc_{i^*} < k_{m_{i^*+1}}$,
by the definition of $i^*$.
Because the labels $m_1, m_2, \ldots$ are in decreasing order of 
server need, $k - \sumsc_{i^*} < k_{m_{i^*}}$.

We will now proceed by enumerating the possible sequences of the $i^*$
largest server needs in $M$.
To prove that $k - \sumsc_{i^*} \le k/6$,
we need only consider such sequences where all server needs are greater than $k/6$.
Such sequences consist only of the elements $k, k/2, k/3, k/4, k/5$.
We enumerate all possible such sequences in \cref{tbl:sequences}.
Note that if $k$ is not divisible by all of $\{2, 3, 4, 5\}$,
some entries will not apply. This only tightens the resulting bound
on $k-\sumsc_{i^*}$ for such $k$.

\begin{table}
\centering
\begin{tabular}{|l|l|l|l|}
\hline
Sequence $k_{m_1}, \ldots, k_{m_{i^*}}$ & $k-\sumsc_{i^*}$
& Sequence $k_{m_1}, \ldots, k_{m_{i^*}}$ & $k-\sumsc_{i^*}$ \\
\hline
$k$ & 0 & $k/2, k/2$ & 0 \\
$k/2,k/3$ & $k/6$ & $k/2,k/4,k/4$ & 0 \\
$k/2,k/4,k/5$ & $k/20$ & $k/2,k/5,k/5$ & $k/10$ \\
$k/3,k/3,k/3$ & 0 & $k/3,k/3,k/4$ & $k/12$ \\
$k/3,k/3,k/5$ & $2k/15$ & $k/3,k/4,k/4$ & $k/6$ \\
$k/3,k/4,k/5,k/5$ & $k/60$ & $k/3,k/5,k/5,k/5$ & $k/15$ \\
$k/4,k/4,k/4,k/4$ & 0 & $k/4,k/4,k/4,k/5$ & $k/20$ \\
$k/4,k/4,k/5,k/5$ & $k/10$ & $k/4,k/5,k/5,k/5$ & $3k/20$ \\
$k/5,k/5,k/5,k/5,k/5$ & 0 & & \\
\hline
\end{tabular}
\caption{All possible sequences of the $i^*$ largest server needs in $M$
in which all server needs exceed $k/6$.}
\label{tbl:sequences}
\end{table}
As shown in \cref{tbl:sequences}, in all cases $k - \sumsc_{i^*} \le k/6$.
The remaining servers are filled with jobs with server need 1.
DivisorFilling-SRPT serves a set of jobs with total server need exactly $k$, as desired.
As a result, \cref{thm:dfs-proof} holds in this case.

\subsubsection{Case 2} Suppose that $k=2^a3^b$ for integers $a, b$,
and that Case 1 does not apply.

Recall that $M_2$ is the set of jobs in $M$ with even server need,
and that $M_3$ is the set of jobs with odd server need,
with server need greater than 1.
We recurse on one of these subsets,
by comparing $2|M_2|$ and $3|M_3|$.
Note that if $M_2$ is recursed on,
the total number of servers in the subproblem is $k/2$,
and all server needs are divisors of $k/2$.
For $M_3$, the same is true of $k/3$.

For \cref{thm:dfs-proof} to hold inductively,
we must show that if $M_2$ is recursed on, then $|M_2| \ge k/2$,
and that if $M_3$ is recursed on, then $|M_3| \ge k/3$.
Because we select a subset by comparing $2|M_2|$ and $3|M_3|$,
if either set is large enough, the set recursed on will be large enough.

Because there are $< n/6$ jobs with server need 1,
$|M_2| + |M_3| \ge 5k/6$.
Therefore, either $|M_2| \ge k/2$ or $|M_3| \ge k/3$.

Suppose that $2|M_2| \ge 3|M_3|$.
Call $M_2'$ the set of jobs in $M_2$, but with all server needs reduced by a factor of 2.
$M_2'$ is the subset that DivisorFilling recurses on.
Because $|M_2'| = |M_2| \ge k/2$ in this case, by our inductive hypothesis
the recursive call returns a subset of $M_2'$ with total server need $k/2$.
The corresponding jobs in $M_2$ have total server need $k$,
so DivisorFilling-SRPT serves a set of jobs with total server need exactly $k$,
completing the inductive step in this case.
If $3|M_3| \ge 2|M_2|$, then $|M_3| \ge k/3$, and the same argument applies.

\subsubsection{Case 3} Suppose that $k$ has largest prime factor $p \ge 5$,
and that Case 1 does not apply.

If $p|M_p| \ge k$, \cref{thm:dfs-proof} holds inductively,
by the same argument as in Case 2.

Let us therefore focus on the extraction procedure.
We must show that the extraction procedure
always extracts $p$ subsets with total server need exactly $k/p$,
to ensure that the overall set served has total server need $k$.

Note that $|M_p| < k/p \le k/5$, and that there are $\le k/6$
jobs with server need 1.
$M_r$ consists of the remaining jobs.
As a result,
\begin{align*}
    |M_r| \ge k - k/6 - k/5 = \frac{19k}{30}.
\end{align*}

Note also that every job in $|M_r|$ has server need at least 2.
The total server need extracted in each step is at most $k/p$,
so the number of jobs extracted is at most $k/2p$.
To prove that $p$ subsets each with total server need $k/p$
can be extracted, it suffices to show that
at least $k/p$ jobs remain after the first $p-1$ subsets have been extracted.

The number of jobs remaining at this point is at least:
\begin{align*}
    \frac{19k}{30} - \frac{(p-1)k}{2p} = \frac{19k}{30} - \frac{k}{2} + \frac{k}{2p}
    = \frac{2k}{15} + \frac{k}{2p}.
\end{align*}
To prove that the number of jobs remaining is at least $k/p$,
we just need to show that $2k/15 \ge k/2p$.
But $p \ge 5$, so $2k/15 > k/10 \ge k/2p$.

Therefore, by induction, each of the $p$ subsets extracted from $M_r$
has total server need $k/p$.
Combining these subsets gives a total server need of $k$.
Therefore, DivisorFilling-SRPT serves a set of jobs with total server need exactly $k$,
as desired.
\end{proof}

\end{document}